\documentclass[12pt]{amsart}
\usepackage{amssymb}
\usepackage{amsmath}
\usepackage{amsfonts}
\usepackage{mathtools}
\usepackage{mathrsfs}
\usepackage{graphicx}
\usepackage{placeins}
\usepackage{pdflscape}
\usepackage{color}
\usepackage[onehalfspacing]{setspace}
\usepackage{caption}
\usepackage{algorithm}
\usepackage{algorithmic}
\usepackage{apptools}
\usepackage{subcaption}
\usepackage{natbib}
\usepackage{enumerate}
\usepackage[utf8]{inputenc}
\usepackage{tikz-cd}
\usepackage[charter,cal=cmcal]{mathdesign}
\usepackage{epigraph}
\usepackage{dcolumn}
\usepackage{placeins}
\usepackage[all,2cell]{xy}
\UseAllTwocells
\xyoption{arc}
\xyoption{rotate}

\usepackage[colorlinks=true,citecolor=blue,urlcolor=blue,pdfpagemode=UseNone,pdfstartview=FitH,hypertexnames=false]{hyperref}

\makeatletter
\renewenvironment{proof}[1][\proofname]{\par
	\normalfont \topsep6\p@\@plus6\p@\relax
	\trivlist
	\item[\hskip\labelsep
		\textbf{#1\@addpunct{:}}]\ignorespaces
}{%
	$\blacksquare$\endtrivlist\@endpefalse
}
\makeatother

\newcommand{\eqdef}{\coloneqq}

\DeclareTextFontCommand{\bi}{%
	\fontseries\bfdefault 
	\itshape
}

\DeclareMathOperator*{\argmin}{\arg\!\min}

\makeatletter
\def\section{\@startsection{section}{1}
	\z@{0.8\linespacing\@plus\linespacing}{.7\linespacing}{\Large}}

\def\subsection{\@startsection{subsection}{2}
	\z@{.5\linespacing\@plus.7\linespacing}{.7\linespacing}{\large}}

\def\subsubsection{\@startsection{subsubsection}{3}
	\z@{.5\linespacing\@plus.7\linespacing}{-.5em}{\normalfont\bfseries}}
\makeatother

\newtheorem{theorem}{Theorem}[section]

\newtheorem{lemma}{Lemma}[section]

\theoremstyle{definition}

\theoremstyle{definition}
\newtheorem{assumption}{Assumption}[section]

\theoremstyle{definition}
\newtheorem{example}{Example}[section]

\calclayout

	\title{}
	
\begin{document}
			\vspace*{5ex minus 1ex}
		\begin{center}
			\Large \textsc{Simple Inference on a Simplex-Valued Weight}
			\bigskip

			\normalsize

		\end{center}
		
		\date{%
			\today%
		}
		
		\vspace*{3ex minus 1ex}
		\begin{center}
		Nathan Canen and Kyungchul Song\\
		\textit{University of Warwick \& CEPR and University of British Columbia}\\
		
		\bigskip
    \bigskip

		\end{center}
		
		\thanks{We thank Bruce Hansen, Frank Schorfheide, and Xiaoxia Shi for conversations that inspired this paper. We also thank Xiaohong Chen, Yiqi Liu, Pujee Tuvaandorj and Yuanyuan Wan, and participants in Canadian Econometrics Study Group Meeting at Carleton University, Cowles Conference in Honor of Don Andrews at Yale University, Econometric Society World Congress in Seoul, and a seminar at University College London for valuable comments. All errors are ours. Song acknowledges that this research was supported by Social Sciences and Humanities Research Council of Canada. Corresponding Address: Kyungchul Song, Vancouver School of Economics, University of British Columbia, 6000 Iona Drive, Vancouver, BC, V6T 1L4, Canada, kysong@mail.ubc.ca}
		\address{Department of Economics, University of Warwick, Coventry, CV4 7AL, United Kingdom}
		\email{nathan.canen@warwick.ac.uk}
		\address{Vancouver School of Economics, University of British Columbia, 6000 Iona Drive, Vancouver, BC, V6T 1L4, Canada}
		\email{kysong@mail.ubc.ca}

		\fontsize{12}{14} \selectfont 	

\begin{abstract}
    In many applications, the parameter of interest involves a simplex-valued weight which is identified as a solution to an optimization problem. Examples include synthetic control methods with group-level weights and various methods of model averaging and forecast combinations. The simplex constraint on the weight poses a challenge in statistical inference due to the constraint potentially binding. In this paper, we propose a simple method of constructing a confidence set for the weight using an adaptive test based on the projection on a polyhedral cone and prove that the method is asymptotically uniformly valid. The procedure does not require tuning parameters or simulations to compute critical values. The confidence set accommodates both the cases of point-identification or set-identification of the weight. We illustrate the method with an empirical example.
\medskip

{\noindent \textsc{Key words:} Simplex-Valued Weight; Synthetic Control; Model Averaging; Forecast Combination; Uniform Asymptotic Validity}
\medskip

{\noindent \textsc{JEL Classification: C30, C54}}
\end{abstract}
\maketitle

\bigskip
\bigskip
\bigskip
\bigskip

\section{Introduction}

We consider the following problem. Suppose that the population weight $w_0 \in \Delta_{K-1}$ in the $(K-1)$-simplex $\Delta_{K-1}$ is defined as a solution to the following optimization problem: 
\begin{align}
	\label{characterization}
	w_0 \in \argmin_{w \in \Delta_{K-1}} \enspace Q(w),
\end{align}
where $Q(w)$ is the population objective function that is convex and differentiable in $w \in \mathbf{R}^{K}$. The dimension $K$ is fixed and not allowed to depend on the sample size $n$. Our focus is on constructing a $(1-\alpha)$-level confidence set $C_{1-\alpha}$ for $w_0$ that is uniformly asymptotically valid.

This inference problem arises in many contexts of applications. For example, in the causal inference literature, various methods of synthetic control design choose a weight as a solution to (\ref{characterization}) (see \cite{Abadie:21:JEL} for a survey of this literature). Meanwhile, in the literature of forecast combinations, the final forecast is constructed as a weighted average of forecasts from different methods or experts. (See \cite{Timmermann:06:Handbook}.) While not in consensus, the use of a simplex-constrained weight has been part of the main approaches considered in this literature. The least squares model averaging proposed by \cite{Hansen:07:Eca} also falls into this framework. In particular, \cite{Hansen:07:Eca} proposed minimizing the Mallows metric to obtain the optimal weight. The population version of the minimization problem can be written as the optimization problem in (\ref{characterization}).
 
The main challenge in the theory of statistical inference on the weight is that the parameter is potentially on the boundary, as $w_0$ can fall on an edge or a vertex of the simplex $\Delta_{K-1}$. Hence, even if $w_0$ is point-identified and $\sqrt{n}$-consistently estimable, it is far from obvious how to construct a pivotal test statistic whose limiting distribution is invariant uniformly over the probabilities in the model. 

Despite the challenge, there are methods which can be applied to develop an asymptotic inference procedure. For example, when $w_0$ satisfies the first order condition for the optimization problem (\ref{characterization}) - an assumption this paper does not assume, we may build a quadratic approximation of the objective function to derive the limiting distribution that depends on the localization parameter (\cite{Geyer:94:AS}, \cite{Andrews:99:Ecma} and \cite{Moon/Schorfheide:09:JOE}), or apply the conditional likelihood ratio approach by \cite{Ketz:18:JOE}. Alternatively, we could adapt the inference based on the random draws from a quasi-posterior in \cite{Chen/Christensen/Tamer:18:Eca} to this setting or employ the bootstrap-based projection method of \cite{Fang/Seo:21:Eca}. These methods target a substantially more general setting than simplex-valued weights. However, they require bootstrap or simulations to obtain critical values and often involve tuning parameters that require a delicate choice to ensure a good finite sample performance.

In this paper, we propose a simple method of constructing a confidence set for the simplex-valued weight which is free of tuning parameters and does not require simulations to compute critical values. Furthermore, our method does not require point-identification of the weight $w_0$ and, hence, does not rely on the quadratic approximation of the objective function. The method is shown to produce confidence sets that are asymptotically valid uniformly over the behavior of the population weight. We provide simulation results and an empirical application to demonstrate the merits of our method. 

Our method relies on a likelihood ratio-type test statistic constrained to a polyhedral cone. When  the test statistic is formed from a multivariate normal random vector, its distribution is known to follow a mixture of a $\chi^2$ distribution with generally unknown weights (see \cite{Silvapulle/Sen:05:ConstrainedStatInference} and references therein). In this setting, \cite{AlMohamad/vanZwet/Cator/Goeman:20:Biometrika} developed an adaptive inference method that selects the relevant $\chi^2$ distribution automatically. In econometrics, \cite{Breunig/Chen:20:WP, Breunig/Chen:24:Eca} proposed a related adaptive approach for testing equality or inequality restrictions on nonparametric functions identified in a nonparametric instrumental variable model. A related idea is also found in \cite{Cox/Shi:23:ReStud} who considered a general moment inequality model with an additive nuisance parameter and proposed asymptotic inference methods. These approaches simplify the inference procedure for testing problems with constraints by using $\chi^2$ distributions with data-dependent degrees of freedom. Our paper builds on this line of research by developing a simple adaptive procedure for asymptotic inference on simplex-valued weights. However, both the specific design of our proposal and the proof of its uniform validity are new, to the best of our knowledge.\footnote{Related to this literature, recently, \cite{Li:25:WP} developed an interesting method of bootstrap-based inference on parameters identified as a constrained optimizer. \cite{Li:25:WP} does not require quadratic approximation of the objective function, yet involves a sequence of tuning parameters that converge to zero. In contrast to \cite{Li:25:WP}, our paper focuses on the inference on a simplex-valued weight, and as such, our proposal is tailored to this case, and does not require point-identification of the weight or tuning parameters that go to zero at a certain rate.}

The rest of the paper proceeds as follows. In Section 2, we introduce a basic set-up with examples, and present our main proposal to construct a confidence set for a simplex-valued weight. In Section 3, we provide our main result of uniform asymptotic validity of the confidence set, and Monte Carlo simulations that study the finite sample properties of our statistical procedure. An empirical application is presented in Section 4 as an illustration. In Section 5, we conclude. Mathematical proofs of the results in this paper are found in the appendix.

\section{A Simple Confidence Set for a Simplex-Valued Weight}

\subsection{A Simplex-Valued Weight}

Let us formally introduce the problem. Suppose that $H_P$ is a $K \times K$ matrix and $h_P$ is a $K$-dimensional vector which potentially depends on the population distribution $P$. Let $\Delta_{K-1}$ denote the simplex in $\mathbf{R}^K$, i.e., $\Delta_{K-1}\eqdef \{w \in \mathbf{R}^K: w \ge 0, w'\mathbf{1} = 1\}$, where $\mathbf{1}$ is the $K \times 1$ vector of ones. (Here, the inequality between two vectors should be understood as the set of pointwise inequalities between the corresponding entries.) 

Given a map $Q_P: \mathbf{R}^K \rightarrow \mathbf{R}$, we define the argmin set of weights, $\mathbb{W}_P$, as follows: 
\begin{align}
    \label{const opt}
	\mathbb{W}_P = \left\{w \in \Delta_{K-1}: Q_P(w) = \inf_{\tilde w \in \Delta_{K-1}} Q_P(\tilde w) \right\}.
\end{align}
We assume that $Q_P$ is convex and differentiable on $\mathbf{R}^K$ and there exists a true weight $w_0$ in the set $\mathbb{W}_P$.

\begin{example}[Synthetic Control with Group-Level Weights]
	\label{exmpl: synthetic control}

	Suppose that we have $K+1$ large groups of individual units which are observed over $T+1$ periods. The units in group $0$ are treated at $T+1$, but all the other units stay untreated. For each time $t=1,...,T+1$, $[Y_{i,t},G_{i,t}]$'s are i.i.d.\ across $i$'s, $Y_{i,t}$ denotes the observed outcome variable and $G_{i,t}$ the group membership taking values in $\{0,1,2,...,K\}$. Let $\mu_{j,t} = \mathbf{E}[Y_{i,t} \mid G_{i,t} = j]$. 
    
    Let $Y_{i,T+1}(1)$ be the potential outcome of an individual $i$ at time $T+1$ when her group $G_{i,T+1}$ is treated first time at time $T+1$ and $Y_{i,T+1}(0)$ that when her group is never treated. We are interested in the average treatment effect on the treated for the target group $0$:
	\begin{align*}
		\theta_0 \eqdef \mathbf{E}[Y_{i,T+1}(1) - Y_{i,T+1}(0) \mid G_{i,T+1} = 0].
	\end{align*}
    For the identification of $\theta_0$, synthetic control with group-level weights replaces a counterfactual untreated average outcome for the target group 0 by a weighted average of donor pool group outcomes. More specifically, synthetic control imposes the following assumption: 
    \begin{align*}
       \mu_{0,T+1}(0) = \sum_{j=1}^K \mu_{j,T+1}(0) w_{0,j},    
    \end{align*}
    where $\mu_{j,T+1}(0) = \mathbf{E}[Y_{i,T+1}(0) \mid G_{i,T+1} = j]$ and $w_0 = [w_{0,1},...,w_{0,K}]'$ is chosen to be such that
	\begin{align*}
		w_0 \in \argmin_{w \in \Delta_{K-1}} \frac{1}{2} \sum_{t=1}^T\left( \mu_{0,t} - \sum_{j=1}^K w_{j} \mu_{j,t} \right)^2.
	\end{align*}
    Then, the target parameter $\theta_0$ is identified as $\theta_0 = \theta_P(w_0) \in \mathbf{R}$, with 
	\begin{align*}
		\theta_P(w) = \mu_{0,T+1} - \sum_{j=1}^K \mu_{j,T+1} w_j. 
	\end{align*}
    
    This setting of a synthetic control method is different from more frequently studied settings where the weights are assigned at the individual level (\cite{Abadie:21:JEL} for a review of this literature.) In contrast, causal inference with groupwise matching assumes large groups of cross-sectional units observed over a short period of time. See \cite{Rincon/Song:25:WP} for a study on this causal inference setting and references therein. 
	
    To see how this example maps to our setting, let $\mu_t = [\mu_{1,t},...,\mu_{K,t}]'$, and define 
	\begin{align*}
		H_P = \frac{1}{T}\sum_{t=1}^T \mu_t \mu_t' \text{ and } h_P = \frac{1}{T}\sum_{t=1}^T \mu_t \mu_{0,t}.
	\end{align*}
	We take
    \begin{align*}
        Q_P(w) = \frac{1}{2} w' H_P w - w'h_P,
    \end{align*}
    and formulate the population optimization problem as follows: 
	\begin{align}
        \label{quadratic programming}
		w_0 \in \argmin_{w \in \Delta_{K-1}} \enspace Q_P(w).
	\end{align}
    As for the weight $w_0$, synthetic control adopts the population-level perfect pretreatment fit condition as follows (\cite{Ferman/Pinto:21:QE}): 
	\begin{align}
		\label{perfect fit}
		\sum_{j=1}^K \mu_{j,t} w_{0,j} = \mu_{0,t}, \text{ for all } t = 1,...,T.
	\end{align}
    In this case, we have $\partial Q_P(w_0)/\partial w = 0$. If $H_P$ is invertible, we have
    \begin{align}
		\label{reg weight}
	   w_0 = H_P^{-1} h_P.
    \end{align}
    Hence, the perfect pre-treatment fit (\ref{perfect fit}) implies that the regression-based weight $w_0 = H_P^{-1} h_P$ falls automatically into the simplex $\Delta_{K-1}$. $\blacksquare$
\end{example}

\begin{example}[Distributional Synthetic Control]
	\label{exmpl: Distr SC}
	\cite{Chen:JAE:2020} and \cite{Gunsilius:23:ECMA} propose methods for constructing counterfactual quantiles of potential outcome distributions using synthetic controls. \cite{Chen:JAE:2020} allows the synthetic control weights to vary across quantiles, whereas \cite{Gunsilius:23:ECMA} employs quantile-invariant weights. In this example, we consider \cite{Gunsilius:23:ECMA}'s approach. Suppose that we have $j = 0, 1, ..., K$ populations, where each population $j$ at time $t$ consists of individuals $i$ with observed outcomes $Y_{i,t}$. Let $Y_{i,t}(0)$ denote the potential outcome of individual $i$ in time $t$ for the untreated state. As in Example \ref{exmpl: synthetic control}, we assume that for each time $t$, the random vector of outcome variable and the group membership, $[Y_{i,t},G_{i,t}]$, are i.i.d.\ across $i$'s. For each $j=0,1,...,K$ and $\tau \in (0,1)$, let $\rho_{j,t}(\tau)$ denote the $\tau$-th quantile of the distribution of $Y_{i,t}$ in population $j$ in time $t$ and $\rho_{j,t}(\tau;0)$ that of $Y_{i,t}(0)$. For simplicity, we assume that we have two periods $t=1,2$, where $t=1$ denotes the pre-treatment period and $t=2$ the post-treatment period. The treatment occurs only in the target population $j=0$. Our focus of interest is the following quantile treatment effect at a given quantile $\tau^*$: 
	\begin{align*}
		\theta_0 \eqdef \rho_{0,2}(\tau^*) - \rho_{0,2}(\tau^*;0),
	\end{align*}
	where $\rho_{0,2}(\tau;0)$ denotes the $\tau$-quantile of $Y_{i,t}(0)$ for individual $i$ in population $0$. 
	
	\cite{Gunsilius:23:ECMA} proposes the following matching condition: 
	\begin{align*}
		\rho_{0,2}(\tau;0) = \sum_{j=1}^K \rho_{j,2}(\tau) w_{j,0},
	\end{align*}
	where $w_0 = [w_{0,1},...,w_{0,K}]' \in \Delta_{K-1}$ is identified by solving the following optimization problem:
    \begin{align*}
        w_0 \in \argmin_{w \in \Delta_{K-1}}\int_0^1 \left( \sum_{j=1}^K \rho_{j,1}(\tau) w_j - \rho_{0,1}(\tau) \right)^2 d\tau.
    \end{align*}
    Then, the target parameter $\theta_0$ is identified as $\theta_P(w)$ with 
	\begin{align*}
		\theta_P(w) = \rho_{0,2}(\tau^*) - \sum_{j=1}^K \rho_{j,2}(\tau^*) w_{j}.
	\end{align*}
	
    Let $U_P(\tau)$ be the $K \times K$ matrix whose $(j,k)$-th entry is given by $\rho_{j,1}(\tau) \rho_{k,1}(\tau)$ and $u_P(\tau)$ the $K$-dimensional vector whose $k$-th entry is given by $\rho_{k,1}(\tau) \rho_{0,1}(\tau)$. Define 
    \begin{align*}
        H_P = \int_0^1 U_P(\tau) d\tau \text{ and } h_P = \int_0^1 u_P(\tau) d\tau.    
    \end{align*} 
    Now, we can reformulate the optimization problem as in (\ref{quadratic programming}) with $Q_P(w) = \frac{1}{2} w' H_P w - w' h_P$. $\blacksquare$
\end{example}

\begin{example}[Forecast Combination]
	In the literature of forecast combination, the final forecast is constructed as a weighted average of multiple forecasts: 
	\begin{align*}
		\hat w \in \argmin_{w \in \Delta_{K-1}} \enspace \frac{1}{n}\sum_{t=1}^n (y_t - \mathbf{\hat y}_t' w)'(y_t - \mathbf{\hat y}_t' w),
	\end{align*} 
	where $y_t$ is the target outcome and $\mathbf{\hat y}_t$ is the $K$ dimensional vector of outcomes used for constructing the forecast at time $t$. (See \cite{Timmermann:06:Handbook} for a review of this literature.) The population version of this optimization problem is given by
	\begin{align*}
		w_0 \in \argmin_{w \in \Delta_{K-1}} \enspace \frac{1}{n}\sum_{t=1}^n \mathbf{E}_P\left[ (y_t - \mathbf{y}_t' w)'(y_t - \mathbf{y}_t' w) \right],
	\end{align*} 
	where $\mathbf{y}_t$ is the $K$ dimensional vector of population outcomes corresponding to $\mathbf{\hat y}_t$. Then, we can rewrite the optimization problem as the minimization of $Q_P(w) = \frac{1}{2} w' H_P w - w'h_P$ over $w \in \Delta_{K-1}$, where 
    \begin{align*}
        H_P = \frac{1}{n}\sum_{t=1}^n \mathbf{E}_P\left[\mathbf{y}_t \mathbf{y}_t'\right] \text{ and } h_P = \frac{1}{n}\sum_{t=1}^n \mathbf{E}_P\left[y_t \mathbf{y}_t\right].
    \end{align*} 
	
	The value of restricting the weight to the simplex has been debated in the literature and defended on the ground that it reduces the variability of the resulting forecast. (See \cite{Timmermann:06:Handbook}.)  Despite the popular use of forecast combinations, the theory of statistical inference on the weighted forecasts appears under-developed (see \cite{Wang/Hyndman/Li/Kang:23:IJF}, p.1539.) $\blacksquare$
\end{example}

\begin{example}[Least Squares Model Averaging]
	\cite{Hansen:07:Eca} proposed a model averaging estimator using the Mallows metric. Let $(y_i,x_i)$, $i=1,...,n$, be a random sample, where $y_i \in \mathbf{R}$ but $x_i = (x_{i1},x_{i2},...)$ is countably infinite. The model for the outcome $y_i$ is given by 
	\begin{align*}
		y_i = \mu_i + e_i,
	\end{align*} 
	where $\mu_i = \sum_{\ell = 1}^\infty a_\ell x_{i \ell}$, $\mathbf{E}[e_i \mid x_i] = 0$, and $a_\ell$'s are constants. Let $0 \le \ell_1 < \ell_2 < ... < \ell_K$, where $\ell_k$'s are integers. For each $k=1,...,K$, we let $\mathbf{a}_k = [a_{1},...,a_{\ell_k}]'$ be the $\ell_k$ dimensional vector of the first $\ell_k$ coefficients. The focus here is on the model average estimator of $\mathbf{a}_K$. Let $\mathbf{\hat a}_k$ be the least squares estimator of $\mathbf{a}_k$ from projecting $Y=[y_1,...,y_n]'$ onto the column space of $X_k$, where $X_k$ denotes the $n \times \ell_k$ matrix whose $(i,j)$-th entry is given by $x_{ij}$. The model averaging estimator $\mathbf{\overline a}_K(w)$ of $\mathbf{a}_K$ takes the following form:
	\begin{align*}
		\mathbf{\overline a}_K(w) \eqdef \hat A w,
	\end{align*} 
	where $\hat A$ is the $\ell_K \times K$ matrix whose $k$-th column vector has the first $\ell_k$ dimensional subvector equal to $\mathbf{\hat a}_k$ and the rest of its entries as zeros. In determining the weight $w = [w_1,...,w_K]'$, \cite{Hansen:07:Eca} proposed to minimize the Mallows metric: 
	\begin{align*}
		\hat w  \in \argmin_{w \in \Delta_{K-1}} (Y - X_K \mathbf{\overline a}_K(w))'(Y - X_K \mathbf{\overline a}_K(w)) + 2 \sigma^2 w' \boldsymbol{\ell},
	\end{align*}
	where $\boldsymbol{\ell} = [\ell_1,...,\ell_K]'$, and showed its asymptotic optimality property. Note that we can write
	\begin{align*}
		\hat w \in \argmin_{w \in \Delta_{K-1}} \enspace \frac{1}{2} w' \hat H w - w' \hat h,
	\end{align*}
	where $\hat H = \frac{1}{n} \hat A' \left( X_K' X_K \right) \hat A$ and $\hat h = \frac{1}{n} \left(\hat A' X_K' Y - \sigma^2 \boldsymbol{\ell} \right)$. Thus, for inference, the population (pseudo-true) weight $w_0$ can be viewed as the solution to the optimization problem (\ref{quadratic programming}), where 
    \begin{align*}
        H_P = \frac{1}{n} A_P' \mathbf{E}_P\left[ X_K' X_K \right]A_P \text{ and } h_P = \frac{1}{n} \left(A_P' \mathbf{E}_P\left[ X_K' Y \right] - \sigma^2 \boldsymbol{\ell} \right),
    \end{align*}
    and $A_P$ is the $\ell_K \times K$ matrix whose $k$-th column vector consists of the first $\ell_k$ dimensional subvector which is equal to $(\mathbf{E}_P[X_k'X_k])^{-1} \mathbf{E}_P[X_k' Y]$ and the rest of its entries as zeros. $\blacksquare$
\end{example}

\subsection{A Simple Confidence Set}
\label{subsec: simple confidence set}

Suppose that the true weight $w_0$ belongs to $\mathbb{W}_P$. Our goal is to construct a confidence set for $w_0$ that is asymptotically uniformly valid over the class of population distributions $P$ under consideration. We will give a set of conditions later. For this section, we present the procedure of constructing the confidence set.

First, we consider the $K \times K$ orthogonal matrix $[\mathbf{1}/\sqrt{K},B_2]$, where $B_2$ is the $K \times (K-1)$ matrix consisting of orthonormal column vectors that are orthogonal to $\mathbf{1}$.\footnote{The computation of $B_2$ is straightforward. First note that $B_2'B_2 = I$. Hence, $B_2 B_2' = I - \mathbf{1} \mathbf{1}'/K$, i.e., the projection matrix projecting onto the orthogonal complement of the span of $\mathbf{1}$. To obtain $B_2$, we obtain a spectral decomposition : $ I - \mathbf{1} \mathbf{1}'/K = U D U'$. From this, we set $B_2$ to be the $K \times (K-1)$ matrix after removing the eigenvector from $U$ that corresponds to the zero diagonal element of $D$.} Define
\begin{align*}
     \varphi_P(w) \eqdef \frac{\partial Q_P(w)}{\partial w},
\end{align*}
and take $\hat \varphi(w)$ as an estimator of $\varphi_P(w)$ such that 
\begin{align*}
	\sqrt{n}B_2'\left(\hat \varphi(w) - \varphi_P(w)\right) \rightarrow_d N(0,V_P(w)),
\end{align*}
as $n \rightarrow \infty$, for some positive definite matrix $V_P(w)$. Suppose that we have a consistent estimator of $V_P(w)$, denoted by $\hat V(w)$. Using $\hat \varphi$ and $\hat V$ only, we can construct a confidence set for $w_0$ as explained in Algorithm \ref{alg:confidence_set}.\footnote{The algorithm applies to cases where $w_0$ is partially identified. When $w_0$ is point-identified and has a consistent estimator $\hat w$, we can substitute $\hat w$ into $\hat V(w)$, replacing $\hat V(w)$ with $\hat V(\hat w)$. This reduces the computation of $\hat \lambda(w)$ to a quadratic programming problem, significantly improving computational speed.} The construction of the confidence set $C_{1-\alpha}$ is simple, involving no simulations  or a sequence of tuning parameters that require a judicious choice in finite samples.

\begin{algorithm}[t]
\caption{Confidence Set for $w_0$}
\label{alg:confidence_set}
\renewcommand{\baselinestretch}{1.5}
\begin{algorithmic}[1]
\REQUIRE $\hat \varphi(w)$ and $\hat{V}(w)$, $w \in \Delta_{K-1}$
\FOR{$w \in \Delta_{K-1}$}
    \STATE \textbf{Step 1:} Solve
    \begin{align*}
    \hat{\lambda}(w) \in \arg\min_{\lambda} \left(\hat \varphi(w) - \lambda\right)' B_2 \hat V^{-1}(w) B_2' \left(\hat \varphi(w) - \lambda\right)
    \end{align*}
    under the constraints: $w'\lambda = 0$ and $\lambda \geq 0$.
    
    \STATE \textbf{Step 2:} Let $\hat{\gamma}(w) = B_2 \hat V^{-1}(w) B_2' \left(\hat \varphi(w) - \hat \lambda(w)\right)$ and 
	\begin{align*}
		\hat d(w) = \left|\left\{j = 1,...,K: \hat{\gamma}_j(w) = 0 \text{ and } w_j = 0 \right\}\right|,
	\end{align*}
	where $\hat \gamma_j(w)$ and $w_j$ denote the $j$-the entry of $\hat \gamma(w)$ and $w$ respectively.
	
    \STATE \textbf{Step 3:} Let $\hat{c}_{1-\alpha}(w)$ be the $1-\alpha$ percentile of the $\chi^2_{\hat{k}(w)}$ distribution, where
    \begin{align*}
    \hat{k}(w) = \max\{K - 1 - \hat{d}(w), 1\}.
    \end{align*}
\ENDFOR
\ENSURE Confidence set:
\begin{align}
C_{1-\alpha} = \left\{w \in \Delta_{K-1}: T(w) \leq \hat{c}_{1-\alpha}(w)\right\},
\end{align}
where 
\begin{align*}
T(w) = n\left(\hat \varphi(w) - \hat \lambda(w)\right)' B_2 \hat V^{-1}(w) B_2' \left(\hat \varphi(w) - \hat \lambda(w)\right).
\end{align*}
\end{algorithmic}
\end{algorithm}

The computation of $V_P(w)$ and its consistent estimation can be done in a standard manner. For example, consider the case where 
\begin{align*}
    Q_P(w) = \frac{1}{2} w' H_P w - w' h_P, \text{ and } \hat Q(w) = \frac{1}{2}w' \hat H w - w' \hat h.
\end{align*}  
Then, we have 
\begin{align*}
    \hat \varphi(w) = \hat H w - \hat h \text{ and } \varphi_P(w) = H_P w - h_P.
\end{align*}
As for $\hat H$ and $\hat h$, suppose that these estimators admit an asymptotic linear representation as follows:
\begin{align*}
	\sqrt{n}(\hat H - H_P) &= \frac{1}{\sqrt{n}}\sum_{i=1}^n \Psi_{i,H} + o_P(1), \text{ and }\\
	\sqrt{n}(\hat h - h_P) &= \frac{1}{\sqrt{n}}\sum_{i=1}^n \psi_{i,h} + o_P(1),
\end{align*}
where $\Psi_{i,H}$ and $\psi_{i,h}$ are influence functions that are i.i.d.\ across $i$'s and have mean zero. Then, we can find that 
\begin{align}
	\label{V_P(w)}
	V_P(w) = \mathbf{E}_P\left[ \xi_i(w) \xi_i(w)'\right],
\end{align}
where $\xi_i(w) = B_2'(\Psi_{i,H} w - \psi_{i,h})$. A consistent estimator $\hat V(w)$ of $V_P(w)$ is obtained as follows: 
\begin{align}
	\label{est V}
	\hat V(w) = \frac{1}{n} \sum_{i=1}^n \hat \xi_i(w) \hat \xi_i(w)',
\end{align}
where $\hat \xi_i(w) = B_2'(\hat \Psi_{i,H} w - \hat \psi_{i,h})$, and $\hat \Psi_{i,H}$ and $\hat \psi_{i,h}$ are appropriate estimators of $\Psi_{i,H}$ and $\psi_{i,h}$. Using $\hat V(w)$, we construct the confidence set $C_{1-\alpha}$ is constructed as explained in Algorithm \ref{alg:confidence_set}.

\begin{example}[Synthetic Control with Group-Level Weights, Cont'd]
    \label{exmpl: synthetic control2}
We revisit Example \ref{exmpl: synthetic control}. Recall that for each $t=1,...,T + 1$, $(Y_{i,t},G_{i,t})$, $i \in N_t$, are i.i.d.\ random vectors. (We allow their distributions to vary over time.) We let $N = \bigcup_{t=1}^{T+1} N_t$ and $n = |N|$. The conditional mean $\mu_{j,t}$ is estimated as a within-group average of the outcomes: for $j=0,1,...,K$ and $t=1,...,T + 1$,
\begin{align*}
    \hat \mu_{j,t} = \frac{1}{n_{j,t}}\sum_{i \in N_{j,t}} Y_{i,t},
\end{align*}
where $N_{j,t} = \{i \in N: G_{i,t} = j\}$. For each group $j$, we have 
\begin{align*}
	\sqrt{n}(\hat \mu_{j,t} - \mu_{j,t}) = \frac{1}{\sqrt{n}}\sum_{i \in N} \psi_{ij,t} + o_P(1),
\end{align*}
where, with $p_{j,t} = P\{G_{i,t} = j\}$,
\begin{align*}
	\psi_{ij,t} = \frac{n}{n_t} \frac{1\{G_{i,t} = j\}}{p_{j,t}}(Y_{i,t} - \mu_{j,t}),
\end{align*}
and $n_t = |N_t|$. If we let $\psi_{i,t} = [\psi_{i1,t},...,\psi_{iK,t}]'$, we have 
\begin{align*}
	\Psi_{i,H} = \frac{1}{T}\sum_{t=1}^T \left(\psi_{i,t} \mu_t' + \mu_t \psi_{i,t}'\right) \text{ and } \psi_{i,h} = \frac{1}{T}\sum_{t=1}^T \left( \mu_t \psi_{i0,t} + \psi_{i,t} \mu_{0,t} \right).
\end{align*}
We obtain $V_P(w) = \mathbf{E}_P\left[ \xi_i(w) \xi_i(w)'\right]$, where $\xi_i(w) = B_2'(\Psi_{i,H} w - \psi_{i,h})$. 

Suppose that the perfect pre-treatment fit (\ref{perfect fit}) is satisfied. In this case, we have 
\begin{align}
	\label{eqs33}
	\Psi_{i,H} w - \psi_{i,h} = \frac{1}{T}\sum_{t=1}^T \mu_t (\psi_{i,t}' w - \psi_{i0,t}) = \frac{1}{T}\sum_{t=1}^T \mu_t \left( \sum_{j=1}^K \psi_{ij,t} w_j - \psi_{i0,t}  \right).
\end{align}
Thus, we obtain the variance formula:
\begin{align}
	\label{VP(w) SC}
	V_P(w) = B_2' \text{Var}_P \left( \frac{1}{T}\sum_{t=1}^T \mu_t \left( \sum_{j=1}^K \psi_{ij,t} w_j - \psi_{i0,t} \right)\right) B_2,
\end{align}
and estimate this by $\hat V(w)$: 
\begin{align*}
	\hat V(w) = B_2' \frac{1}{n}\sum_{i \in N} \left( \frac{1}{T}\sum_{t=1}^T \hat \mu_t \left( \sum_{j=1}^K \hat \psi_{ij,t} w_j - \hat \psi_{i0,t} \right)\right) \left( \frac{1}{T}\sum_{t=1}^T \hat \mu_t \left( \sum_{j=1}^K \hat \psi_{ij,t} w_j - \hat \psi_{i0,t} \right)\right)' B_2,
\end{align*} 
where, with $\hat p_{j,t} = n_{j,t}/n_t$,
\begin{align*}
	\hat \psi_{ij,t} = \frac{n}{n_t} \frac{1\{G_{i,t} = j\}}{\hat p_{j,t}}(Y_{i,t} - \hat \mu_{j,t}).
\end{align*}

When the sample is repeated cross-sections, the observations are independent across time. In this case, we can obtain sharper inference by modifying $\hat V(w)$ as follows: 
\begin{align}
	\label{VP(w) SC2}
	\hat V(w) = \frac{1}{n}\sum_{i \in N} \frac{1}{T^2}\sum_{t=1}^T \left( \sum_{j=1}^K \hat \psi_{ij,t} w_j - \hat \psi_{i0,t} \right)^2 B_2' \hat \mu_t \hat \mu_t' B_2.
\end{align}
$\blacksquare$
\end{example}

\begin{example}[Distributional Synthetic Control, Cont'd]
    \label{exmpl: Distr SC Contd}
We revisit Example \ref{exmpl: Distr SC}. Let $\hat\rho_{j,t}(\tau)$ be the empirical $\tau$-quantile of $Y_{i,t}$, $i \in N_{j,t}$ for $j = 0,1,...,K$, where $N_{j,t} = \{i \in N: G_{i,t} = j\}$. For brevity, we assume that there exist constants $c,C>0$ such that for all $\tau \in (0,1)$, $f_{j,t}(\rho_{j,t}(\tau)) \in (c,C)$, where $f_{j,t}$ is the density of $Y_{i,t}$, $i \in N_{j,t}$, in time $t$.  Then, we can show that
\begin{align*}
	\sqrt{n}(\hat \rho_{j,t}(\tau) - \rho_{j,t}(\tau)) = \frac{1}{\sqrt{n}}\sum_{i \in N} \psi_{ij,t}(\tau) + o_P(1),
\end{align*}
where the $o_P(1)$ term is uniform over $\tau \in (0,1)$, and 
\begin{align}
	\label{psi_ik,t}
	\psi_{ij,t}(\tau) = \frac{n 1\{i \in N_{j,t}\}}{n_{j,t}} \cdot \frac{\tau - 1\{Y_{i,t} \le \rho_{j,t}(\tau)\}}{f_{j,t}(\rho_{j,t}(\tau))}.
\end{align}
(See e.g. Section 2.5 of \cite{Serfling:80:Approx}.\footnote{The uniformity of $o_P(1)$ in $\tau \in (0,1)$ can be shown using the empirical process theory in the standard way.}) Then, the $(j,k)$-th entry of $\Psi_{i,H}$ is given as follows: 
\begin{align*}
	\int_0^1 \left( \psi_{ij,1}(\tau) \rho_{k,1}(\tau) + \psi_{ik,1}(\tau) \rho_{j,1}(\tau) \right)d\tau,
\end{align*}
and the $k$-th entry of $\psi_{i,h}$ is given by 
\begin{align*}
	\int_0^1 \left( \psi_{i0,1}(\tau) \rho_{k,1}(\tau) + \psi_{ik,1}(\tau) \rho_{0,1}(\tau) \right)d\tau.
\end{align*}
We obtain $V_P(w) = \mathbf{E}_P\left[ \xi_i(w) \xi_i(w)'\right]$, where $\xi_i(w) = B_2'(\Psi_{i,H} w - \psi_{i,h})$. $\blacksquare$
\end{example}

\begin{algorithm}[t]
\caption{Confidence Set for $w_0$ with a Bootstrap Variance Estimator}
\renewcommand{\baselinestretch}{1.5}
\label{alg:bootstrap_confidence_set}
\begin{algorithmic}[1]
\REQUIRE $\hat \varphi$ and $\hat{V}^*(\hat{w})$
\FOR{$w \in \Delta_{K-1}$}
    \STATE \textbf{Step 1:} Solve
    \begin{align*}
    \tilde{\lambda}(w) &\in \arg\min_{\lambda} \left(\hat \varphi(w) - \lambda\right)' B_2 \hat V^{* - 1}(\hat{w}) B_2' \left(\hat \varphi(w) - \lambda\right)
    \end{align*}
    under the constraints: $w'\lambda = 0$ and $\lambda \geq 0$.
    
    \STATE \textbf{Step 2:} Let $\tilde{\gamma}(w) = B_2 \hat V^{* - 1}(\hat{w}) B_2' \left(\hat \varphi(w) - \hat \lambda(w)\right)$ and 
	\begin{align*}
		\tilde d(w) = \left|\left\{j = 1,...,K: \tilde{\gamma}_j(w) = 0 \text{ and } w_j = 0 \right\}\right|,
	\end{align*}
	where $\tilde \gamma_j(w)$ and $w_j$ denote the $j$-the entry of $\hat \gamma(w)$ and $w$ respectively.
    
    \STATE \textbf{Step 3:} Let $\tilde{c}_{1-\alpha}(w)$ be the $1-\alpha$ percentile of the $\chi^2_{\tilde{k}(w)}$ distribution, where
    \begin{align*}
    \tilde{k}(w) = \max\{K - 1 - \tilde{d}(w), 1\}.
    \end{align*}
\ENDFOR
\ENSURE Confidence set:
\begin{align}
C^*_{1-\alpha} = \left\{w \in \Delta_{K-1}: \tilde{T}(w) \leq \tilde{c}_{1-\alpha}(w)\right\},
\end{align}
where 
\begin{align*}
\tilde{T}(w) = n\left(\hat \varphi(w) - \tilde \lambda(w)\right)' B_2 \hat V^{* - 1}(\hat{w}) B_2' \left(\hat \varphi(w) - \tilde \lambda(w)\right).
\end{align*}
\end{algorithmic}
\end{algorithm}

In some applications, the computation and estimation of $V_P(w)$ can be cumbersome for practitioners, because the practitioner has to find an analytical form of $V_P(w)$ in their application to find its consistent estimator. In such cases, an alternative is to use a bootstrap estimator of $V_P(w)$. As shown by \cite{Hahn/Liao:21:Ecma}, the asymptotic validity of the inference procedure is preserved under the bootstrap, while the inference procedure is conservative in general.\footnote{In some settings, we can modify the bootstrap variance estimator using a truncation method proposed by \cite{Shao:92:Stat} to render the inference procedure asymptotically non-conservative. See also \cite{Goncalves/White:05:JASA}.} Suppose that $\hat \varphi^*(w)$ is constructed from the bootstrap sample, such that the bootstrap distribution of $\sqrt{n}B_2'(\hat \varphi^*(w) - \hat \varphi(w))$ converges almost surely to $N(0,V_P(w))$. Suppose that $w_0$ is point-identified and the minimizer $\hat w$ of $\hat Q(w)$ over $w \in \Delta_{K-1}$ is consistent. Then, we obtain the bootstrap variance estimator:
\begin{align*}
	\hat V^*(\hat w) = B_2' \mathbf{E}^*\left[ n (\hat \varphi^*(\hat w) - \hat \varphi(\hat w))(\hat \varphi^*(\hat w) - \hat \varphi(\hat w))' \right] B_2,
\end{align*}
where $\mathbf{E}^*$ denotes the expectation with respect to the bootstrap distribution.  This suggests the modified inference procedure in Algorithm \ref{alg:bootstrap_confidence_set}. Note that $\hat V^*(\hat w)$ is outside the iteration over $w \in \Delta_{K-1}$. Hence, once the bootstrap estimator $\hat V^*(\hat w)$ is computed, the cost of the computation remains the same as before.

\subsection{Heuristics}

\subsubsection{Construction of the Test Statistic}

Let us give a heuristic motivation behind the construction of the confidence set in Algorithm \ref{alg:confidence_set}. To construct a test statistic, we form the Lagrangian of the constrained optimization in (\ref{const opt}):
\begin{align*}
	\mathcal{L}(w,\tilde \lambda, \lambda) = Q_P(w) + \tilde \lambda (w^{\prime} \mathbf{1}-1) - \lambda^{\prime} w,
\end{align*}
where $\tilde \lambda$ and $\lambda$ are Lagrange multipliers. The necessary and sufficient condition for $w_0$ to be a solution to the constrained optimization is: 
\begin{align*}
    \varphi_P(w_0) + \tilde \lambda \mathbf{1} - \lambda = 0,
\end{align*}
for some $\tilde \lambda \in \mathbf{R}$ and $\lambda \in U(w_0)$, where 
\begin{align*}
    U(w_0) = \left\{ \lambda \in \mathbf{R}^K: w_0' \lambda = 0 \text{ and } \lambda \ge 0 \right\}.
\end{align*} 
We concentrate out $\tilde \lambda$ to obtain the following equality: 
\begin{align*}
    (I - \mathbf{1}\mathbf{1}'/K)\left( \varphi_P(w_0) - \lambda \right) = 0,
\end{align*}
for some $\lambda \in U(w_0)$. The equality has $K$ equations but these equations are not linearly independent. To extract maximally linearly independent equations out of these, we premultiply $B_2'$ to obtain the following equality restriction:
\begin{align}
	\label{equality restr}
    B_2'\left( \varphi_P(w_0) - \lambda \right) = 0,
\end{align}
for some $\lambda \in U(w_0)$. The equality restrictions motivate the test statistic $T(w)$ in Algorithm \ref{alg:confidence_set}.

\subsubsection{Construction of a Critical Value}

To provide an intuition behind the construction of the critical value, it helps to begin with the result of \cite{AlMohamad/vanZwet/Cator/Goeman:20:Biometrika}. Suppose that $Y \sim N(\mu,V)$, for some $\mu \in \mathbf{R}^K$ and a symmetric positive definite matrix $V$. We introduce a polyhedral cone $\Lambda(w)$ as follows: 
\begin{align}
    \label{Lambda Omega}
	\Lambda(w) := \left\{ B_2' \lambda: w'\lambda = 0 \text{ and } \lambda \ge 0, \lambda \in \mathbf{R}^{K} \right\}.
\end{align}
Define the norm $\| \cdot \|_{V}$ as $\|x\|_{V} = \sqrt{x' V^{-1} x}$, $x \in \mathbf{R}^K$. Then we focus on the test statistic of the form: 
\begin{align*}
    T_0(w) = \left\| Y - \Pi_{V}(Y \mid \Lambda(w)) \right\|_V^2,
\end{align*}
where $\Pi_{V}(Y \mid C)$ for a closed convex set $C$ denotes the projection of $Y$ on $C$ along $\| \cdot \|_{V}$. Let $\Lambda^\circ(w)$ be the polar cone of $\Lambda(w)$ and let $F_{\ell}$, $\ell=1,...,L$, be the faces of $\Lambda^\circ(w)$ and $\text{ri}(F_\ell)$ their relative interiors. Furthermore, let $k_\ell$ be the rank of the projection matrix of $Y$ onto the linear span of $F_{\ell}$. Then, Theorem 1 of \cite{AlMohamad/vanZwet/Cator/Goeman:20:Biometrika} gives the following result: 
\begin{align*}
    P\{\|Y - \Pi_{V}(Y \mid \Lambda(w))\|^2 > q_{1-\alpha,V}(Y;w) \} \le \alpha,
\end{align*}
where\footnote{\cite{AlMohamad/vanZwet/Cator/Goeman:20:Biometrika} did not consider the case where the projection of $Y$ onto the polar cone becomes zero with positive probability. Thus, here we take $\max\{k_\ell,1\}$ in place of $k_\ell$ as in their paper.} 
\begin{align}
    \label{critical value}
    q_{1-\alpha,V}(Y;w) = \sum_{\ell = 1}^L G^{-1}(1-\alpha;\max\{k_\ell,1\})1\{\Pi_{V}(Y \mid \Lambda^\circ(w)) \in \text{ri}(F_\ell)\},
\end{align} 
and $G(\cdot;k)$ denotes the CDF of the $\chi^2$ distribution with $k$ degrees of freedom. In order to adapt their result to our setting, we take two steps. First, we find a characterization of the critical value in our setting. Second, we extend the validity result to the desired asymptotic validity result. 

We illustrate the main idea here, assuming that $V = I$ and is known. We write simply $q_{1-\alpha}(Y;w)$ instead of $q_{1-\alpha,V}(Y;w)$ and $\Pi(Y \mid C)$ instead of $\Pi_{V}(Y \mid C)$. The general case where $V$ is unknown and consistently estimated is dealt with in the proof of the main result in the appendix.\medskip 

\textbf{Step 1 (Critical Value Characterization):} We first establish the following characterization of the critical value $q_{1-\alpha}(Y;w)$:
\begin{align*}
    q_{1-\alpha}(Y;w) = G^{-1}(1-\alpha;k(w)),
\end{align*}
where
\begin{align*}
	k(w) &:= \max\{K-1 - d(w),1\} \text{ and } \\
	d(w) &:= |J_0[B_2 (Y - \Pi(Y \mid \Lambda(w)))] \cap J_0[w]|,
\end{align*} 
and $J_0[x] = \{j: x_j = 0\}$, the set of the indices of zeros in a vector $x$. To show this, we find the polar cone $\Lambda^\circ(w)$ of $\Lambda(w)$ as
\begin{align*}
    \Lambda^\circ(w) = \left\{ x \in \mathbf{R}^{K-1}: [B_{2}x]_{J_0[w]} \le 0 \right\},
\end{align*}
where for any vector $a$, $[a]_J$ denotes the subvector of $a$ indexed by $J$. From this, the faces of $\Lambda^\circ(w)$ and their relative interiors are seen to be of the following form: for $J \subset J_0[w]$,
\begin{align*}
    \Lambda_{J}^\circ(w) &= \left\{ x \in \mathbf{R}^{K-1}: [B_{2} x]_J = 0 \text{ and } [B_{2} x]_{J_0(w) \setminus J} \le 0 \right\}, \text{ and }\\
    \text{ri}(\Lambda_{J}^\circ(w)) &= \left\{ x \in \mathbf{R}^{K-1}: [B_{2} x]_J = 0 \text{ and } [B_2 x]_{J_0[w] \setminus J} < 0 \right\}.
\end{align*}
The linear span of $\text{ri}(\Lambda_{J}^\circ(w))$ is $L_J^\circ := \{x \in \mathbf{R}^{K-1}: [B_2 x]_J = 0\}$. Hence, the rank of the projection matrix onto this space is $K-1 - |J|$. Furthermore, since the relative interiors, $\text{ri}(\Lambda_{J}^\circ(w))$, partition the polyhedral cone $\Lambda^\circ(w)$, we find that 
\begin{align}
	\label{J0 = J}
    \Pi(Y \mid \Lambda^\circ(w)) \in \text{ri}(\Lambda_{J}^\circ(w)) \text{ if and only if } J_0[B_2 \Pi(Y \mid \Lambda^\circ(w))] \cap J_0[w] = J.
\end{align} 
The projection of $Y$ onto $\Lambda^\circ(w)$ falls on $\text{ri}(\Lambda_{J}^\circ(w))$ if and only if the projection falls on the latter's linear span $L_J^\circ$. Therefore, only one of the terms in the sum on the right-hand side of (\ref{critical value}) realizes. Noting the decomposition 
\begin{align*}
	Y = \Pi(Y \mid \Lambda(w)) + \Pi(Y \mid \Lambda^\circ(w)),
\end{align*}
we write this term as $G^{-1}(1-\alpha;k(w))$.

\begin{figure}[t]
	\begin{center}
		\includegraphics[scale=0.48]{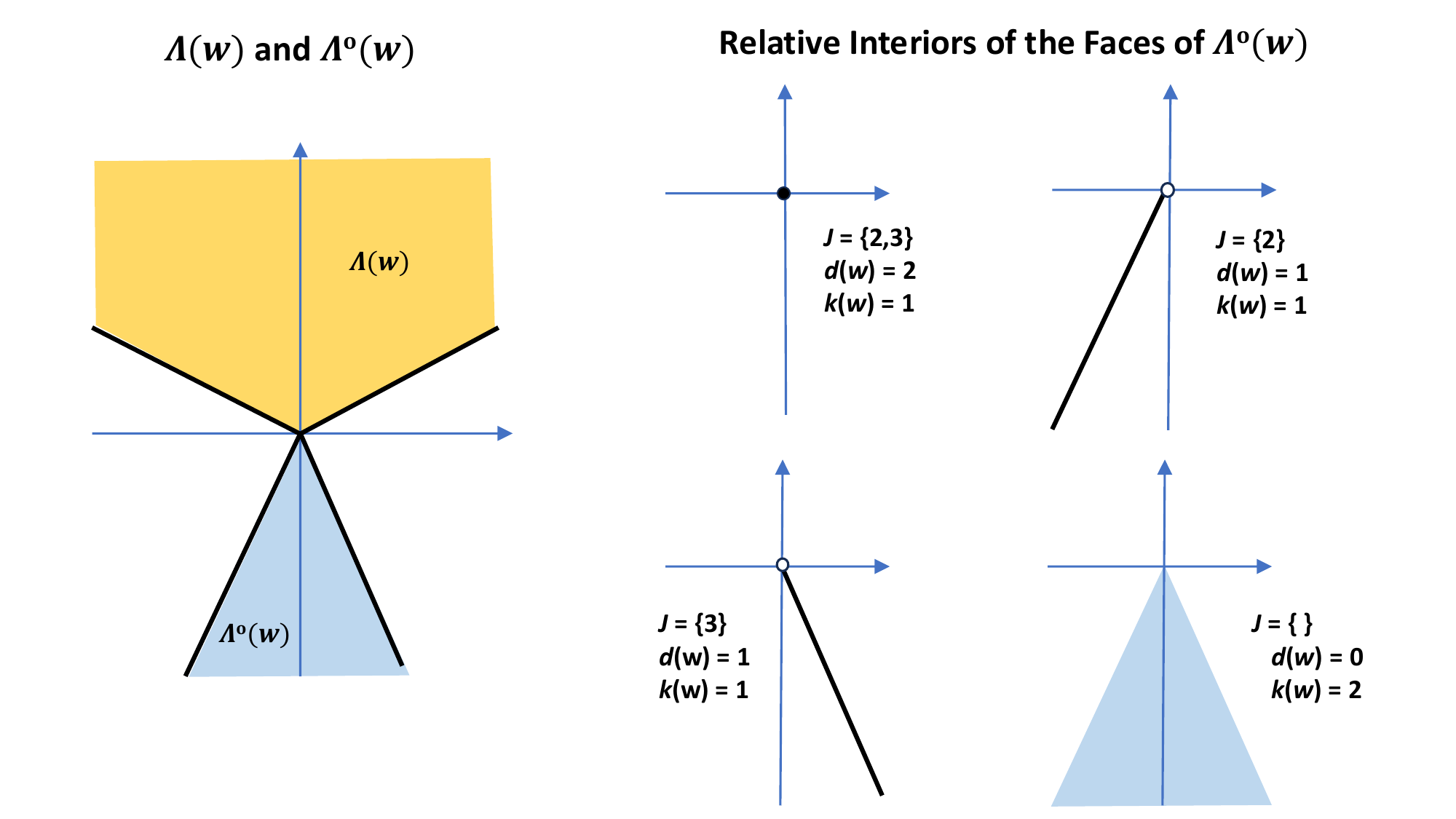}
        \captionsetup{width=1\textwidth}
		\caption{The Relative Interiors of the Faces of $\Lambda^\circ(w)$ and the Degrees of Freedom for $\chi^2$ Distribution}
    
		\label{fig:PolarCones}

        \parbox{\textwidth}{\footnotesize Note: The left plot depicts $\Lambda(w)$ and $\Lambda^\circ(w)$ for the case of $K=3$ and $w = [1,0,0]'$. The four plots on the right depicts the relative interiors of the four faces of $\Lambda^\circ(w)$, presented with the corresponding $d(w)$ and $k(w)$. The relative interiors partition $\Lambda^\circ(w)$. The degrees of the $\chi^2$ distribution depend on which relative interior the projection of $Y$ onto $\Lambda^\circ(w)$ falls. For example, the relative interior with $J = \{2,3\}$ is $\{0\}$ with the projection reduced to zero. In that case, the degree of freedom of the $\chi^2$ distribution is taken to be 1. The degrees of freedom are largest ($k(w) = 2$), with probability one, when the projection falls on the relative interior of the face $\Lambda_{J}^\circ(w)$ with $J = \varnothing$.
    }
	\end{center}
\end{figure}

To illustrate the degrees of freedom $k(w)$ for the $\chi^2$ distribution in the critical value, consider the case $K=3$ and $w = [1,0,0]'$. In this case, the cone $\Lambda(w)$ and its polar cone $\Lambda^\circ(w)$ are given as follows: 
\begin{align*}
	\Lambda(w) &= \{(x_1,x_2) \in \mathbf{R}^2 : x_2 \ge x_1/\sqrt{3}, x_2\ge -x_1/\sqrt{3}\}, \text{ and }\\
	\Lambda^\circ(w) &= \{(x_1,x_2) \in \mathbf{R}^2 : x_2 \le \sqrt{3} x_1, x_2 \le -\sqrt{3} x_1\}.
\end{align*} 
There are four faces of $\Lambda^\circ(w)$: $\Lambda_{\{2,3\}}^\circ(w)$, $\Lambda_{\{2\}}^\circ(w)$, $\Lambda_{\{3\}}^\circ(w)$, and $\Lambda_{\varnothing}^\circ(w)$. The relative interiors of these faces and their corresponding degrees of the $\chi^2$ distribution in the critical value are depicted in Figure \ref{fig:PolarCones}.\medskip

\textbf{Step 2 (Extension to Asymptotic Validity): } So far the result is for a normally distributed random vector $Y$. In order to accommodate random vectors that are asymptotically normal, we extend the previous result. This extension consists of two components: the asymptotic approximation of the test statistic and that of the critical values. For simplicity, we maintain the assumption that $\hat V_n$ is known to be $I$. 

Consider a setting where 
\begin{align*}
	Y_n = Z_n + \mu_n, \text{ with } Z_n \rightarrow_d Z \sim N(0,I),
\end{align*}
as $n \rightarrow \infty$. (Note that we allow $\|\mu_n\| \rightarrow \infty$ as $n \rightarrow \infty$.) From Skorohod representation, there exists a probability space and a sequence of random vectors $\tilde Z_n$ that have the same distribution as $Z_n$ and $\tilde Z_n \rightarrow_{a.s.} Z$ for a random vector $Z \sim N(0,I)$. We let $\tilde Y_n = \tilde Z_n + \mu_n$ and $Y_n^* = Z + \mu_n$. We focus on the limit behavior of the following test statistic: 
\begin{align*}
	\tilde T_n(w) = \left\| \tilde Y_n - \Pi(\tilde Y_n \mid \Lambda(w)) \right\|^2, \enspace w \in \Delta_{K-1}.
\end{align*}
Since the projection map on a nonempty, closed convex set is a contraction map, we have 
\begin{align*}
	\left\|\Pi(\tilde Y_n \mid \Lambda^\circ(w)) - \Pi(Y_n^* \mid \Lambda^\circ(w)) \right\| \le \| \tilde Y_n - Y_n^* \| \rightarrow_{a.s.} 0,
\end{align*}
as $n \rightarrow \infty$. Hence, it is not hard to see that, with $T_n^*(w) :=  \left\|Y_n^* - \Pi(Y_n^* \mid \Lambda(w)) \right\|^2$, 
\begin{align*}
	\tilde T_n(w) - T_n^*(w) \rightarrow_{a.s.} 0, \text{ as } n \rightarrow \infty.
\end{align*}

The main challenge is to show the asymptotic validity of the critical values. For this, we prove that for any subsequence of $\{n\}$, there exists a further subsequence $\{n'\} \subset \{n\}$, with large probability,
\begin{align}
	\label{convergence}
	J_0[B_2\Pi(\tilde Y_n \mid \Lambda^\circ(w))]  \cap J_0[w] \subset J_0[B_2\Pi(Y_n^* \mid \Lambda^\circ(w))]  \cap J_0[w].
\end{align}
This yields the result that the critical value $c_n^*$ based on $Y_n^*$ is less than the critical value $\tilde c_n$ based on $\tilde Y_n$ eventually. Hence, the rejection probability $P\{T_n^*(w) > c_n^*\}$ is bounded below by $P\{\tilde T_n(w) > \tilde c_n\}$. By the construction of the critical value $c_n^*$ using the $\chi^2$ distribution, the former rejection probability is bounded by $\alpha$, delivering the asymptotic validity of the test.   

It remains to show (\ref{convergence}). Since $|J_0[x]|$ is discontinuous in $x$, this result does not come directly from the convergence, $\tilde Y_n - Y_n^* \rightarrow_{a.s.} 0$. We begin by choosing any subsequence and find a further subsequence, along with some $J,J' \subset J_0[w]$ such that
\begin{align*}
	\Pi(\tilde Y_n \mid \Lambda^\circ(w)) \in \text{ri}(\Lambda_{J}^\circ(w)) \text{ and } \Pi(Y_n^* \mid \Lambda^\circ(w)) \in \text{ri}(\Lambda_{J'}^\circ(w)).
\end{align*}
We show that the set of values of $\tilde Y_n$ such that $J$ is not contained in $J'$ eventually has measure zero. In light of (\ref{J0 = J}), this yields (\ref{convergence}). While this summarizes the basic insights briefly, delivering the proof considering random variance matrices $\hat \Omega$ and the sequences of weights $w_n \in \Delta_{K-1}$ requires a considerable care for subtleties in the proofs. We refer the reader to the appendix for details.

\subsection{Discussions}

In this section, we discuss related methods developed in the literature and compare them with ours. For brevity, we discuss two methods that are most closely related to ours. First, we discuss the method of quadratic approximation of the sample objective function as proposed by \cite{Geyer:94:AS} and \cite{Andrews:99:Ecma} and a related proposal by \cite{Ketz:18:JOE}. Second, we discuss \cite{Cox/Shi:23:ReStud} who proposed inverting a test whose critical values are based on a $\chi^2$ distribution with a data dependent degree of freedom. Like our approach, their procedure does not involve tuning parameters that require a judicious choice to ensure stable finite sample properties.

\subsubsection{Quadratic Approximation}

Our sample objective function $\hat Q(w)$ is a quadratic function of $w$. Hence, we may consider applying the method of quadratic approximation to address the issue of $w_0$ on the boundary of the simplex $\Delta_{K-1}$. To see how this method applies to our setting, consider the example of synthetic control with groupwise matching which takes the sample objective function $\hat Q(w)$ of the form: 
\begin{align*}
    \hat Q(w) = \frac{1}{2} w' \hat H w - w' \hat h,
\end{align*}
where $\hat H$ and $\hat h$ are $\sqrt{n}$-consistent and asymptotically normal estimators of a positive definite matrix $H$ and a vector $h$. Now, we can write 
\begin{align}
	\label{quad approx}
    \hat Q(w) - \hat Q(w_0) = (w - w_0)'(\hat H w_0 - \hat h) + \frac{1}{2}(w - w_0)' \hat H (w - w_0),
\end{align}
for $w_0 \in \argmin_{w \in \Delta_{K-1}} Q(w)$. The quadratic approximation method requires 
\begin{align*}
    \sqrt{n}(\hat H w_0 - \hat h) \to_d \zeta, \text{ as } n \to \infty,
\end{align*}
for some random variable $\zeta$. Under regularity conditions, this requirement is met if 
\begin{align}
	\label{eq restr}
	H w_0 - h = 0.
\end{align}
More generally, the quadratic approximation requires the following: at $w = w_0$,
\begin{align*}
    \frac{\partial Q_{P}(w)}{\partial w} = 0.
\end{align*}
This means that the global minimizer of $Q_P$ lies in the simplex $\Delta_{K-1}$. However, when the simplex constraint is binding, we do not have this equality and cannot the methods based on the quadratic approximation. The setting of synthetic control naturally uses additional equality restrictions (\ref{perfect fit}) which represent perfect pre-treatment fit at the population level. In this case, the condition (\ref{eq restr}) is satisfied, and we can apply the method of quadratic approximation. However, as it is well known, due to the constraint imposed on the estimator $\hat w$, the limiting distribution of $\sqrt{n}(\hat w - w)$ is characterized as a minimizer of a random function. Thus, in order to obtain the critical value, we need to first simulate the random function and obtain a minimizer. In contrast to these approaches, our method is simple, as it does not require simulating the limiting distribution of the test statistic. More importantly, our method does not require that the objective function admit a quadratic approximation or the parameter $w_0$ be point-identified.

\cite{Ketz:18:JOE} proposed a simple, interesting way to deal with a parameter on the boundary when the sample objective function admits a quadratic approximation. The estimator is obtained by applying a single Newton-Raphson adjustment to the constrained estimator. Let us study his approach in the context of the synthetic control setting. Let $\hat w$ be the constrained estimator of $w_0$. Then, from the quadratic approximation in (\ref{quad approx}), his approach results in the following estimator: 
\begin{align*}
	\tilde w = \hat w - \hat H^{-1} (\hat H \hat w - \hat h) = \hat H^{-1} \hat h.
\end{align*}
This is the regression estimator by regressing $\hat \mu_{0,t}$ onto $\hat \mu_{1,t}$,...,$\hat \mu_{K,t}$. Hence, the estimator $\tilde w$ is an unconstrained estimator, allowed to take values outside $\Delta_{K-1}$. Due to this, the estimator is $\sqrt{n}$ consistent and asymptotically normal, under a regularity conditions where $w_0$ is $\sqrt{n}$-consistently estimable. 

To map the construction of this test statistic to our setting, note that using the pre-treatment fit condition in synthetic control: $H w_0 = h$, we can write
\begin{align*}
	\sqrt{n}B_2' \hat H (\tilde w - w_0) &= - B_2' (\sqrt{n}(\hat H - H) w_0 - \sqrt{n}(\hat h - h)) \\
	&= - \sqrt{n}\hat f(w_0), \text{ say}.
\end{align*}
This motivates the following quasi-likelihood ratio test statistic:
\begin{align*}
	T_{\mathsf{NR}}(w) \eqdef n \hat f(w)' \hat V^{-1}(w) \hat f(w).  
\end{align*}
Under regularity conditions, we expect that $T_{\mathsf{NR}}(w_0) \to_d \chi_{K-1}^2$, as $n \to \infty$. Thus, the resulting method is \textit{non-adaptive} to the possibility of $w_0$ being on the boundary.\footnote{\cite{Hsieh/Shi/Shum:22:JoE} proposed a method of statistical inference on equality or inequality restrictions using a linear constraint formulation via Karush-Kuhn-Tucker conditions. We can adapt their proposal to this setting of simplex-valued weights. Like ours, their main proposal does not require a tuning parameter or simulating critical values. However, when adapted to our setting, the critical values are taken from the $\chi^2$ distribution with a degree of freedom $K+1$ (see (39) on page 258 of their paper.) Thus, the test is more conservative than the non-adaptive version above.} In contrast, our method is based on the idea from \cite{AlMohamad/vanZwet/Cator/Goeman:20:Biometrika} and adaptive to $w_0$ being on the boundary. For comparison, our test statistic takes the form: 
\begin{align*}
	T(w) = n (\hat f(w) - B_2' \hat \lambda(w))' \hat V^{-1}(w) (\hat f(w) - B_2' \hat \lambda(w)),
\end{align*}
and the critical values are taken from $\chi_{\hat k(w)}^2$. While there is no uniform dominance of one test over the other, \cite{AlMohamad/vanZwet/Cator/Goeman:20:Biometrika} present the power comparison between the two approaches and report simulation results in favor of the adaptive test.

\subsubsection{Adaptive Moment Inequality Tests}

Closely related to our proposal is that of \cite{Cox/Shi:23:ReStud} who proposed an adaptive size-exact subvector inference from moment inequality restrictions. They considered the moment inequality model:\footnote{Importantly, \cite{Cox/Shi:23:ReStud}'s framework accommodates a model with a nuisance parameter that enters the moment inequality restrictions in an additive form. To facilitate the comparison, we only discuss a simplified version of their model without a nuisance parameter here.} 
\begin{align*}
	A \mathbf{E}_P[\overline m_n(\theta)] \le b,
\end{align*}
where $A$ is a $d_A \times d_m$ matrix, $b$ is a $d_A$ dimensional vector, and $\overline m_n(\theta)$ is the sample average of the moment function $m(W_i,\theta)$:
\begin{align*}
	\overline m_n(\theta) = \frac{1}{n}\sum_{i=1}^n m(W_i,\theta).
\end{align*}
To construct a confidence set, they considered the following test statistic: 
\begin{align}
	\label{Cox/Shi}
 T_n(\theta) = \inf_{\mu: A\mu \le b} n (\overline m_n(\theta) - \mu)'\hat \Sigma(\theta)^{-1}(\overline m_n(\theta) - \mu),
\end{align}
where
\begin{align*}
	\hat \Sigma(\theta) = \frac{1}{n}\sum_{i=1}^n (m(W_i,\theta) - \overline m_n(\theta))(m(W_i,\theta) - \overline m_n(\theta))'.
\end{align*}
Like our proposal, they constructed the critical value from a $\chi^2$ distribution with data-dependent degrees of freedom. 

While our problem appears similar to theirs, their methods and asymptotic validity results do not subsume ours. First, our testing framework is not necessarily motivated from moment inequality restrictions. Hence, we cannot directly use their uniform asymptotic validity result without modification in our setting. Second, our method is simpler than theirs in our case. To see this, we rewrite $T(w)$ as follows:
\begin{align*}
	T(w) = \inf_{\tilde \lambda \in \Lambda(w)} n \left(\hat f(w) - \tilde \lambda \right)^{\prime} \hat V^{-1}(w) \left( \hat f(w) - \tilde \lambda \right),
\end{align*}
where $\Lambda(w)$ is the polyhedron defined in (\ref{Lambda Omega}) and $\hat f(w) = B_2'\hat \varphi(w)$. To the best of our knowledge, there does not seem to be an explicit solution for a matrix $A(w)$ and a vector $b(w)$ such that 
\begin{align*}
	\Lambda(w) = \left\{ \lambda \in \mathbf{R}^{K-1}: A(w) \lambda \le b(w)\right\}.
\end{align*}
While one may develop an algorithm to compute $A(w)$ and $b(w)$ and apply the approach of \cite{Cox/Shi:23:ReStud}, our proposal is already simpler than this, as we do not need to compute $A(w)$ and $b(w)$ for each $w \in \Delta_{K-1}$.

Alternatively, we may consider rewriting $T(w)$ as follows:
\begin{align*}
	T(w) = \inf_{\lambda: w'\lambda = 0, \lambda \ge 0} n \left(\hat \varphi(w) - \lambda \right)^{\prime} B_2 \hat V(w)^{-1} B_2' \left( \hat \varphi(w) - \lambda \right),
\end{align*}
to map this to the setting of \cite{Cox/Shi:23:ReStud}. In this case, we can explicitly find matrices $A(w)$ and a vector $b(w)$ such that 
\begin{align*}
	\left\{\lambda \in \mathbf{R}^K : w'\lambda = 0, \lambda \ge 0 \right\} = \left\{ \lambda \in \mathbf{R}^K: A(w) \lambda \le b(w)\right\}.
\end{align*}
 However, unlike $\hat \Sigma(\theta)$ in (\ref{Cox/Shi}), the matrix $B_2 \hat V(w)^{-1} B_2'$ is a singular matrix both in finite samples and in the limit. Hence, the results and proofs of \cite{Cox/Shi:23:ReStud} are not directly applicable here. The singularity problem in our setting stems from the linear constraint $w'\mathbf{1} = 1$ rather than from accommodating distributions with singular covariance matrices in uniformly asymptotically valid inference. Therefore, simply removing such distributions from consideration cannot resolve this issue.

\subsection{Extension: Inference on a Function of $w_0$}
\label{subsec: extension}
In many applications, the main parameter of interest is an identified function of $w_0$: 
\begin{align*}
	\theta_0 = \theta_P(w_0),
\end{align*}
for some map $\theta_P: \Delta_{K-1} \to \Theta$ and $\Theta \subset \mathbf{R}^L$ is the parameter space for $\theta_0$.\footnote{Recently, \cite{Dufour/Tuvaandorj:25:arXiv} developed a quasi-likelihood ratio test when the parameter of interest is a known function of a nuisance parameter that is on the boundary, and showed its pointwise asymptotic validity. Their test is simple to use, involving no tuning parameters or bootstrap for critical values. \cite{VanDijcke/Gunsilius/Wright:24:arXiv} developed bootstrap inference for treatment effect parameters identified via the distributional synthetic control approach of \cite{Gunsilius:23:ECMA} and proved its pointwise asymptotic validity. However, their proof requires Hadamard differentiability of the parameter functional at distributions outside a certain exceptional set, thereby precluding uniform asymptotic validity over any region containing this set.} In this setting, we consider two approaches for constructing a confidence set for $\theta_0$. 

The first approach simply uses the Bonferroni method. More specifically, suppose that the estimator $\hat \theta$ satisfies the following asymptotic linear representation: 
\begin{align}
	\label{theta influence}
	\sqrt{n}(\hat \theta(w_0) - \theta(w_0)) &= \frac{1}{\sqrt{n}}\sum_{i=1}^{n} \psi_{i,\theta}(w_0) + o_P(1) \\ \notag
	&\quad \quad \rightarrow_d N(0,\Sigma_P(w_0)), \enspace \text{ as } n \to \infty,
\end{align}
where $\Sigma_P(w_0)$ is a symmetric positive definite matrix, and $n$ denotes the size of the sample from the population that identifies $\theta_P(\cdot)$. Then, for any $\alpha \in (0,1)$, we can use the confidence set $C_{1-\kappa}$ for $w_0$ with level $100(1-\kappa)\%$, $\kappa \in (0,\alpha)$ (in Algorithm \ref{alg:confidence_set}) and construct the $100(1-\alpha)\%$-level confidence interval for $\theta$ as 
\begin{align}
    \label{CI theta}
	\tilde C_{1-\alpha}^{\mathsf{Bonf}} \eqdef \left\{\theta \in \Theta : \inf_{w \in C_{1-\kappa}} n (\hat \theta(w) - \theta)' \hat \Sigma^{-1}(w) (\hat \theta(w) - \theta) \le q_{1 - \alpha - \kappa} \right\},
\end{align}
where $\hat \Sigma(w)$ is a consistent estimator of $\Sigma_P(w)$ and $q_{1-\alpha - \kappa}$ denotes the $(1-\alpha-\kappa)$ quantile of the $\chi^2$ distribution with degree of freedom $L$.

The second approach is to construct a joint confidence set for $[w_0',\theta_P(w_0)']'$ and project the confidence set on the space for $\theta_P(w_0)$.\footnote{We are grateful to a referee from our previous submission for suggesting this idea.} First, let us assume that for each $w \in \Delta_{K-1}$,
\begin{align}
	\label{asym joint normal}
	\sqrt{n} \begin{bmatrix}
	    B_2'(\hat \varphi(w) - \varphi_P(w))\\
	   \hat \theta(w) - \theta_P(w)
	\end{bmatrix}
    \to_d N\left(0, \Omega_P(w) \right),
\end{align}
as $n \to \infty$, for a symmetric positive definite covariance matrix $\Omega_P(w)$, where $\hat \varphi(\cdot)$ is a $\sqrt{n}$-consistent estimator of $\varphi_P(\cdot)$. (Note that we allow for the sample size for $\hat \theta(\cdot)$ to be different from that for $\hat \varphi(\cdot)$.) We let 
\begin{align*}
	\hat s(w,\theta,\lambda) = 
	\begin{bmatrix}
	   \hat \varphi(w) - \lambda\\
	   \hat \theta(w) - \theta
	\end{bmatrix}
	\text{ and }
	\tilde B_2 = \begin{bmatrix}
	         B_2 & 0_{K \times L}\\
			0_{L \times (K-1)} & I        
	\end{bmatrix}
	.
\end{align*}
Then, using these and a consistent estimator $\hat \Omega(w)$ of $\Omega_P(w)$, we construct a confidence set $C_{1-\alpha}^{\mathsf{joint}}$ as in Algorithm \ref{alg:confidence_set3}. The confidence set for $\theta$ can be obtained from projecting the joint confidence set onto $\Theta$: 
\begin{align}
	\label{CI theta2}
	\tilde C_{1-\alpha}^{\mathsf{proj}} = \left\{\theta \in \Theta: (w,\theta) \in C_{1-\alpha}^{\mathsf{joint}}, \text{ for some } w \in \Delta_{K-1} \right\}.
\end{align}
While the projection method can be among various methods of subvector inference, we have opted for this method due to its simplicity, being free of any tuning parameters. We relegate the investigation of improvement on this method to future research.

\begin{algorithm}[t]
\caption{Joint Confidence Set for $(w_0,\theta_0)$}
\label{alg:confidence_set3}
\renewcommand{\baselinestretch}{1.5}
\begin{algorithmic}[1]
\REQUIRE $\hat s(w,\theta,\lambda)$ and $\hat \Omega(w)$, $\lambda \in \mathbf{R}^K$, $w \in \Delta_{K-1}$ and $\theta \in \mathbf{R}^L$.
\FOR{$(w,\theta) \in \Delta_{K-1} \times \Theta$}
    \STATE \textbf{Step 1:} Solve
    \begin{align*}
    \hat{\lambda}(w,\theta) &\in \arg\min_{\lambda} \enspace \hat s(w,\theta,\lambda)' \tilde B_2 \hat \Omega^{-1}(w) \tilde B_2' \hat s(w,\theta,\lambda)
    \end{align*}
    under the constraints: $w'\lambda = 0$ and $\lambda \geq 0$.
    
    \STATE \textbf{Step 2:} Let $\hat{\gamma}(w,\theta) = B \hat \Omega^{-1}(w) \tilde B_2' \hat s(w,\theta,\hat \lambda(w,\theta))$ and 
	\begin{align*}
		\hat d(w,\theta) = \left|\left\{j = 1,...,K: \hat{\gamma}_j(w,\theta) = 0 \text{ and } w_j = 0 \right\}\right|,
	\end{align*}
	where $\hat \gamma_j(w,\theta)$ denotes the $j$-the entry of $\hat \gamma(w,\theta)$, and
\begin{align}
	\label{B}
	B \eqdef [ B_2,0_{K \times L}].
\end{align}
	 
    \STATE \textbf{Step 3:} Let $\hat{c}_{1-\alpha}(w,\theta)$ be the $1-\alpha$ quantile of the $\chi^2_{\hat{k}(w,\theta)}$ distribution, where
    \begin{align*}
    \hat{k}(w,\theta) = \max\{K + L - 1 - \hat{d}(w,\theta), 1\}.
    \end{align*}
\ENDFOR
\ENSURE Joint confidence set:
\begin{align}
C_{1-\alpha}^{\mathsf{joint}} = \left\{(w,\theta) \in \Delta_{K-1} \times \Theta : T(w,\theta) \leq \hat{c}_{1-\alpha}(w,\theta)\right\},
\end{align}
where 
\begin{align*}
T(w,\theta) = n \times \hat s(w,\theta,\hat \lambda(w,\theta))' \tilde B_2 \hat \Omega^{-1}(w) \tilde B_2' \hat s(w,\theta,\hat \lambda(w,\theta)).
\end{align*}
\end{algorithmic}
\end{algorithm}

\begin{example}[Synthetic Control with Group-Level Weights, Cont'd]
    \label{exmpl: synthetic control30}
	
    Let us revisit the synthetic control setting in Example \ref{exmpl: synthetic control}. For an estimator of $\theta_P(w)$, we consider $\hat \theta(w) = \hat \mu_{0,T+1} - \hat \mu_{T+1}' w.$ Recall that 
	\begin{align*}
		\sqrt{n} B_2' (\hat \varphi(w) - \varphi_P(w)) = \frac{1}{\sqrt{n}} \sum_{i \in N} \frac{1}{T}\sum_{t=1}^T B_2'\mu_t \left( \sum_{j=1}^K \psi_{ij,t} w_j - \psi_{i0,t} \right) + o_P(1),
	\end{align*}
	where $\psi_{ij,t}$, $j=0,1,...,T$, are defined in Example \ref{exmpl: synthetic control2}. Also, note that
	\begin{align*}
		\sqrt{n}(\hat \theta(w) - \theta_P(w)) = \frac{1}{\sqrt{n}} \sum_{i \in N} \left( \psi_{i0,T+1} - \sum_{j=1}^K \psi_{ij,T+1} w_j \right) + o_P(1),
	\end{align*}
	where $\hat \theta(w) = \hat \mu_{0,T+1} - \sum_{j=1}^K \hat \mu_{j,T+1} w_j.$ When we apply the Bonferroni method, we can construct 
	\begin{align*}
		\hat \sigma^2(w) = \frac{1}{n} \sum_{i \in N} \left( \hat \psi_{i0,T+1} - \sum_{j=1}^K \hat \psi_{ij,T+1} w_j \right)^2,
	\end{align*}
	with 
	\begin{align*}
		\hat \psi_{ij,T+1} = \frac{n}{n_{T+1}} \frac{1\{G_{i,T+1} = j\}(Y_{i,T+1} - \hat \mu_{j,T+1})}{\hat p_{j,T+1}},
	\end{align*}
	where $\hat p_{j,T+1} = n_{j,T+1}/n_{T+1}$. Thus, we obtain 
	\begin{align}
    \label{CI theta Bonf}
	\tilde C_{1-\alpha}^{\mathsf{Bonf}} \eqdef \left\{\theta \in \Theta : \inf_{w \in C_{1-\kappa}} \frac{n (\hat \theta(w) - \theta)^2}{\hat \sigma^2(w)} \le q_{1 - \alpha + \kappa} \right\},
    \end{align}
	with $q_{1 - \alpha + \kappa}$ defined as before, except with $L=1$. 

	As for the projection method, we construct 
	\begin{align}
		\label{hat Omega(w)}
		\hat \Omega(w) &= \begin{bmatrix}
		     \hat \Omega_{11}(w) & \hat \Omega_{12}(w) \\
			 \hat \Omega_{21}(w) & \hat \Omega_{22}(w)
	\end{bmatrix}
	,
	\end{align}
	where 
	\begin{align*}
		\hat \Omega_{11}(w) &= \frac{1}{n} \sum_{i \in N} B_2' \left( \frac{1}{T}\sum_{t=1}^T \hat \mu_t \left(\sum_{j=1}^K \hat \psi_{ij,t} w_j - \hat \psi_{i0,t}\right) \right) \left( \frac{1}{T}\sum_{t=1}^T \hat \mu_t \left(\sum_{j=1}^K \hat \psi_{ij,t} w_j - \hat \psi_{i0,t}\right) \right)' B_2\\
		\hat \Omega_{12}(w) &= \frac{1}{n} \sum_{i \in N} B_2' \left( \frac{1}{T}\sum_{t=1}^T \hat \mu_t \left(\sum_{j=1}^K \hat \psi_{ij,t} w_j - \hat \psi_{i0,t}\right) \right) \left( \hat \psi_{i0,T+1}  - \sum_{j=1}^K \hat \psi_{ij,T+1} w_j \right), \text{ and } \\
		\hat \Omega_{22}(w) &= \frac{1}{n} \sum_{i \in N} \left(\hat \psi_{i0,T+1} - \sum_{j=1}^K \hat \psi_{ij,T+1} w_j \right)^2.
	\end{align*}
	
	When the sample is repeated cross-sections, we can utilize independence across time and obtain sharper joint inference for $(w_0,\theta_0)$, by setting
	\begin{align*}
		\hat \Omega(w) = \begin{bmatrix}
                         \hat \Omega_{11}(w) & 0\\
                         0 & \hat \Omega_{22}(w)		
                  		\end{bmatrix},
	\end{align*}
	where 
	\begin{align*}
		\hat \Omega_{11}(w) = \frac{1}{n} \sum_{i \in N} \frac{1}{T^2}\sum_{t=1}^T \left(\sum_{j=1}^K w_j \hat \psi_{ij,t} - \hat \psi_{i0,t} \right)^2 B_2' \hat \mu_t \hat \mu_t' B_2,
	\end{align*}
	and $\hat \Omega_{22}(w)$ is the same as defined above. Note that $\Omega_{12}(w)$ is zero, because it consists of covariances between quantities involving observations at $t=1,...,T$ and those involving observations at $t = T+1$. 

	Now, using $\hat \Omega(w)$, we obtain the joint confidence set $C_{1-\alpha}^{\mathsf{joint}}$ as in Algorithm \ref{alg:confidence_set3}. From this, we obtain the confidence set for $\theta_0$ as in (\ref{CI theta2}). $\blacksquare$
\end{example}

\begin{example}[Distributional Synthetic Control, Cont'd]
	\label{exmpl: Distr SC Contd2}
    Let us revisit Example \ref{exmpl: Distr SC Contd}. In this setting, we have 
	\begin{align*}
		\sqrt{n} B_2' (\hat \varphi(w) - \varphi_P(w)) = \frac{1}{\sqrt{n}} \sum_{i \in N} B_2'(\Psi_{i,H} w - \psi_{i,h}) + o_P(1),
	\end{align*}
	where $\Psi_{i,H}$ and $\psi_{i,h}$ are defined in Example \ref{exmpl: Distr SC Contd}. Define 
	\begin{align*}
		\hat \theta(w) = \hat \rho_{0,2}(\tau^*) - \sum_{j=1}^K \hat \rho_{j,2}(\tau^*) w_j.
	\end{align*}
	As for the estimated quantile treatment effect,
	\begin{align*}
		\sqrt{n}(\hat \theta(w) - \theta_P(w)) = \frac{1}{\sqrt{n}} \sum_{i \in N} \left( \psi_{i0,2}(\tau^*) - \sum_{j=1}^K \psi_{ij,2}(\tau^*) w_j \right) + o_P(1),
	\end{align*}
	where $\psi_{ij,t}(\tau)$ is defined in (\ref{psi_ik,t}). When we apply the Bonferroni method, we can construct 
	\begin{align*}
		\hat \sigma^2(w) = \frac{1}{n} \sum_{i \in N} \left( \hat \psi_{i0,2}(\tau^*) - \sum_{j=1}^K \hat \psi_{ij,2}(\tau^*) w_j \right)^2,
	\end{align*}
	with 
	\begin{align*}
		\hat \psi_{ij,2}(\tau^*) = \frac{n}{n_{j,2}} \frac{1\{i \in N_{j,2}\}(\tau^* - 1\{Y_{i,2} \le \hat \rho_{j,2}(\tau^*)\})}{\hat f_{j,2}(\hat \rho_{j,2}(\tau^*))},
	\end{align*}
	where $\hat f_{j,2}(t)$ is a consistent and asymptotically normal estimator of $f_{j,2}(t)$. Using $\hat \theta(w)$ and $\hat \sigma^2(w)$, we can construct the confidence set using the Bonferroni method as before.

	As for the projection method, we construct 
	\begin{align}
		\label{hat Omega(w)2}
		\hat \Omega(w) &= \begin{bmatrix}
		     \hat \Omega_{11}(w) & \hat \Omega_{12}(w) \\
			 \Omega_{21}(w) & \Omega_{22}(w)
	\end{bmatrix}
	,
	\end{align}
	where 
	\begin{align*}
		\hat \Omega_{11}(w) &= \frac{1}{n} \sum_{i \in N} B_2' (\hat \Psi_{i,H} w - \hat \psi_{i,h})(\hat \Psi_{i,H} w - \hat \psi_{i,h})' B_2\\
		\hat \Omega_{12}(w) &= \frac{1}{n} \sum_{i \in N} B_2' (\hat \Psi_{i,H} w - \hat \psi_{i,h})\left( \hat \psi_{i0,2}(\tau^*) - \sum_{j=1}^K \hat \psi_{ij,2}(\tau^*) w_j \right), \text{ and } \\
		\hat \Omega_{22}(w) &= \frac{1}{n} \sum_{i \in N} \left( \hat \psi_{i0,2}(\tau^*) - \sum_{j=1}^K \hat \psi_{ij,2}(\tau^*) w_j \right)^2,
	\end{align*}
	and $\hat \Psi_{i,H}$ and $\hat \psi_{i,h}$ are estimated versions of $\Psi_{i,H}$ and $\psi_{i,h}$ with $\psi_{ij,t}$ replaced by $\hat \psi_{ij,t}$. Using $\hat \Omega(w)$, we obtain the joint confidence set $C_{1-\alpha}^{\mathsf{joint}}$ as in Algorithm \ref{alg:confidence_set3}.
\end{example}

\section{Uniform Asymptotic Validity}
\label{subsec: uniform validity}

\subsection{A Preliminary Result}

We first provide a preliminary result. Suppose that $Y_n \in \mathbf{R}^{K+L-1}$, $n \ge 1$, is a sequence of random vectors such that 
    \begin{align}
        \label{Yn}
        Y_n = \hat \Omega^{1/2} Z_n + \mu_n,
    \end{align} 
where $Z_n \in \mathbf{R}^{K+L-1}$ is a sequence of asymptotically standard normal random vectors, $\hat \Omega$ is a sequence of symmetric positive definite $(K+L-1) \times (K+L-1)$ random matrices, and $\mu_n \in \mathbf{R}^{K+L-1}$ is a sequence of nonstochastic vectors. We do not assume that the sequence $\mu_n$ is bounded.

Our main focus is on the test statistic that is based on the squared error from the projection of $Y_n$ onto a polyhedral cone along the norm $\|\cdot \|_{\hat \Omega}$, where $\| x \|_{\hat \Omega}^2 = x' \hat \Omega^{-1} x$, $x \in \mathbf{R}^{K+L-1}$, and $\Lambda(w)$ is defined in (\ref{Lambda Omega}). Recall that $G(\cdot;k)$ denotes the CDF of the $\chi^2$ distribution with the degrees of freedom equal to $k$ and $J_0[z] = \{1\le j \le K: z_j = 0\}$ for any vector $z = [z_j] \in \mathbf{R}^K$. For each $w \in \Delta_{K-1}$, we define
\begin{align}
	\label{tilde Lambda(w)}
    \tilde \Lambda(w) = \left\{\begin{bmatrix}
                              B_2' \lambda\\
                              0_{L \times 1}
                            \end{bmatrix}
                             \in \mathbf{R}^{K + L - 1}: \lambda \ge 0, w'\lambda = 0 \right\}.
\end{align}
The test statistic takes the form of the squared residuals from projecting $Y_n$ onto $\tilde \Lambda(w)$: 
\begin{align*}
	\left\| Y_n - \Pi_{\hat \Omega}(Y_n\mid \tilde \Lambda(w)) \right\|_{\hat \Omega}^2,
\end{align*}
where $\Pi_{\hat \Omega}(Y_n\mid \tilde \Lambda(w))$ denotes the projection of $Y_n$ onto $\tilde \Lambda(w)$ along $\| \cdot \|_{\hat \Omega}$. 

As for the critical value, let $\hat \delta(w) \eqdef B \hat \Omega^{-1} (Y_n - \Pi_{\hat \Omega}(Y_n \mid \tilde \Lambda(w)))$ and 
\begin{align*}
	\hat k(w) \eqdef \max\left\{K+L-1 - |J_0[\hat \delta(w)] \cap J_0[w]|,1 \right\},
\end{align*}
with $B$ defined in (\ref{B}) and $0_{K \times L}$ denoting the $K \times L$ dimensional matrix of zeros. Then, the construction of the confidence set for $w$ is based on the critical value of the following form: 
\begin{align*}
	\hat c_{1-\alpha}(Y_n;w) := G^{-1}\left(1-\alpha; \hat k(w) \right).
\end{align*}
This critical value is random, depending on the data through its dependence on the number of zeros in $\hat \delta(w)$.

We first introduce an assumption that requires the consistency of $\hat \Omega$ and the asymptotic normality of $Z_n$ along a sequence of probabilities $P_n$.
\begin{assumption}
    \label{basic assumption}
    For each $t \in \mathbf{R}$ and $\epsilon>0$, 
    \begin{align*}
        P_n\left\{Z_n \le t \right\} \to \Phi(t) \text{ and } P_n\left\{ \left\|\hat \Omega - \Omega_n \right\| > \epsilon \right\} \rightarrow 0,
    \end{align*}
    for some sequence of nonstochastic $(K+L-1) \times (K+L-1)$ symmetric positive definite matrices $\Omega_n$ such that for some $\overline B, \epsilon >0$, $\| \Omega_n\| \le \overline B$ and $\lambda_{\min}(\Omega_n) \ge \epsilon$, for all $n \ge 1$, where $\lambda_{\min}(\Omega_n)$ denotes the minimum eigenvalue of $\Omega_n$ and $\Phi$ the CDF of $N(0,I)$.
\end{assumption}

The asymptotic normality of $Z_n$ and the consistency of $\hat \Omega$ are typically satisfied in the applications under consideration in this paper. The following lemma is crucial for the uniform asymptotic validity result. 

\begin{lemma}
	\label{lemm: Mohamad2}
	Suppose that Assumption \ref{basic assumption} holds. Then, for any $\alpha \in (0,1)$, and for any sequence $w_n \in \Delta_{K-1}$ and $\mu_n \in \tilde \Lambda(w_n)$, we have
	\begin{align*}
		\limsup_{n \rightarrow \infty} P_n\left\{\left\| Y_n - \Pi_{\hat \Omega}(Y_n\mid \tilde \Lambda(w_n)) \right\|_{\hat \Omega}^2 > \hat c_{1-\alpha}(Y_n;w_n)\right\} \le \alpha.
	\end{align*}
\end{lemma}

This lemma shows that reading the critical values from the $\chi^2$ distribution with the degrees of freedom equal to $\hat k(w)$ generates a uniformly asymptotically valid procedure. The proof of the lemma can be potentially useful for many similar settings where the test statistic involves a projection onto a polyhedral cone. 

\subsection{The Main Result}

Let us revisit Section \ref{subsec: extension} and present the main result of the asymptotic validity of the confidence set for $(w_0,\theta_0)$. We focus on the uniform asymptotic validity of the joint confidence set $C_{1-\alpha}^{\mathsf{joint}}$ in Algorithm \ref{alg:confidence_set3}, as it is more general than $C_{1-\alpha}$ in Algorithm \ref{alg:confidence_set}. First, we introduce a set of assumptions. From here on, $\mathcal{P}_n$ denotes the set of probability distributions of data that are under consideration. Our goal is to establish the uniform asymptotic validity of the confidence set $C_{1-\alpha}^{\mathsf{joint}}$ for $w_0$ over the class of probability distributions $\mathcal{P}_n$.

\begin{assumption}
	\label{assump: asym normal}

	For each sequence of probabilities $P_n \in \mathcal{P}_n$ and each $w \in \Delta_{K-1}$, there exists a sequence of symmetric positive definite matrices, $\Omega_{P_n}(w)$, such that the following conditions are satisfied for any sequence of weights $w_n \in \mathbb{W}_{P_n}$.\medskip 

   (i) For each $t \in \mathbf{R}^{K + L -1}$, we have
   \begin{align*}
	P_n\left\{ \Omega_{P_n}^{-1/2}(w_n) \sqrt{n}
		\begin{bmatrix}
			B_2'(\hat \varphi(w_n) - \varphi_{P_n}(w_n)) \\
            \hat \theta(w_n) - \theta_{P_n}(w_n)	 
	    \end{bmatrix} 
		\le t \right\} \rightarrow \Phi(t),
   \end{align*}
   as $n \rightarrow \infty$, where $\Phi$ is the CDF of the standard normal random vector.

   (ii) $\limsup_{n \rightarrow \infty} \left\|\Omega_{P_n}(w_n) \right\| < \infty$ and $\liminf_{n \rightarrow \infty}\lambda_{\min}(\Omega_{P_n}(w_n)) > 0$.

   (iii) For any $\epsilon>0$, and along any sequence $w_n \in \Delta_{K-1}$, 
   \begin{align*}
	 P_{n}\left\{ \left\|\hat \Omega(w_n) - \Omega_{P_n}(w_n) \right\| > \epsilon \right\} \rightarrow 0.
   \end{align*}
\end{assumption}
\medskip

Condition (i) requires the asymptotic normality of the estimators that is uniform over the probabilities in $\mathcal{P}_n$. This condition is satisfied in many settings where maps, $B_2' \varphi_P(w)$, $\theta_P(w)$, are ``regular'', i.e., behave ``smoothly'' in the perturbation of $P$. Condition (ii) requires that the limit distribution is non-degenerate uniformly over the probabilities in $\mathcal{P}_n$. Condition (iii) requires the variance estimator $\hat \Omega(w_n)$ to be consistent for $\Omega_P(w_n)$ uniformly over the probabilities in $\mathcal{P}_n$. Condition (iv) requires that the minimizer of $Q_{P_n}(w)$ over $w \in \Delta_{K-1}$ is close to the global minimizer. This condition is easily satisfied with the synthetic control examples with the pretreatment fit. In these cases, we have $\varphi_{P_n}(w) = 0$ for all $w \in \mathbb{W}_{P_n}$. The constant $C$ is permitted to be unknown, and hence the condition (iv) is too weak to apply the approach of quadratic approximation. 

Thus, the conditions in Assumption \ref{assump: asym normal} essentially define the scope of the method and are satisfied in many settings. By applying Lemma \ref{lemm: Mohamad2}, we obtain the following result.
\begin{theorem}
	\label{thm: main} 
	Suppose that Assumption \ref{assump: asym normal} holds. Then, for any $\alpha \in (0,1)$, 
	\begin{align*}
		\liminf_{n \rightarrow \infty} \inf_{P \in \mathcal{P}_n} \inf_{w \in \mathbb{W}_P} P\left\{ (w,\theta_{P}(w)) \in C_{1-\alpha}^{\mathsf{joint}}\right\} \ge 1 - \alpha.
	\end{align*}
\end{theorem}

The proof is found in the appendix of this paper. To see how Lemma \ref{lemm: Mohamad2} gives this validity result, observe that 
\begin{align*}
	T(w,\theta) = \left\| Y_n(w,\theta)- \Pi_{\hat \Omega(w)}(Y_n(w,\theta) \mid \tilde \Lambda(w)) \right\|_{\hat \Omega(w)}^2,
\end{align*}
where 
\begin{align}
	\label{Yn(w)}
	Y_n(w,\theta) = \sqrt{n} \begin{bmatrix}
			B_2' \hat \varphi(w) \\
            \hat \theta(w) - \theta	 
	    \end{bmatrix}.
\end{align}
Thus, the test statistic $T(w,\theta)$ is nothing but the squared length of the projection error in projecting $Y_n(w)$ onto the polyhedral cone $\Lambda(w)$ along the norm $\| \cdot \|_{\hat \Omega(w)}$. Since 
\begin{align*}
	Y_n(w,\theta_{P_n}(w)) = \hat \Omega^{1/2}(w) Z_n(w) + \mu_n(w),
\end{align*}
with 
\begin{align}
	\label{Zn(w)}
	Z_n(w) =  \sqrt{n} \hat \Omega^{-1/2}(w) \begin{bmatrix}
			B_2' (\hat \varphi(w) - \varphi_{P_n}(w))\\
            \hat \theta(w) - \theta_{P_n}(w)	 
	    \end{bmatrix}
		\text{ and } \mu_n(w) = \sqrt{n}
		\begin{bmatrix}
			B_2' \varphi_{P_n}(w) \\
            0	 
	    \end{bmatrix},
\end{align}
the random vector $Y_n(w,\theta_{P_n}(w))$ is asymptotically normal, yet with a potentially diverging shift $\mu_n(w)$. This setting maps to the one in Lemma \ref{lemm: Mohamad2}.\footnote{Note that we allow for the possibility that $\liminf_{n \rightarrow \infty}\varphi_{P_n}(w_n) \ne 0$ for $w_n \in \mathbb{W}_{P_n}$ such as when the minimizer of $Q_{P_n}(w)$ over $w \in \mathbf{R}^K$ lies outside $\Delta_{K-1}$. In this case, the sequence $\|\mu_n(w_n)\|$ may diverge to $\infty$, as $n \rightarrow \infty$.} The uniform asymptotic validity of the confidence set $C_{1-\alpha}$ for $w_0$ follows from the lemma.

\begin{example}[Synthetic Control with Group-Level Weights, Cont'd]
    \label{exmpl: synthetic control3}
	
    Let us consider low level conditions for Assumption \ref{assump: asym normal} for the synthetic control setting in Example \ref{exmpl: synthetic control30}. It is not hard to see that Assumption \ref{assump: asym normal} is satisfied if the following conditions are satisfied.\medskip
	
	(i) For each $t=1,...,T$, the random vectors, $[Y_{i,t},G_{i,t}]$, $i \in N_t$, are i.i.d., and 
	\begin{align*}
		\limsup_{n \ge 1} \sup_{P \in \mathcal{P}_n} \max_{0 \le j \le K} \mathbf{E}_P\left[ Y_{i,t}^4 \mid G_{i,t} = j \right] < \infty \text{ and } \liminf_{n \ge 1} \inf_{P \in \mathcal{P}_n} \min_{0 \le j \le K} P\left\{G_{i,t} = j \right\} > 0.
	\end{align*}

	(ii) $\liminf_{n \ge 1} \inf_{P \in \mathcal{P}_n} \inf_{w \in \Delta_{K-1}} \lambda_{\min}\left(\Omega_P(w)\right)>0$, where $\Omega_P(w)$ denotes the population version of $\hat \Omega(w)$ defined in (\ref{hat Omega(w)}).\medskip

    Assumption \ref{assump: asym normal}(i) is satisfied by the Central Limit Theorem and Law of Large Numbers. Note that Assumption \ref{assump: asym normal}(ii) is violated if $Y_{i,t}$ is identically distributed across time $t$. For example, in the case of a linear factor model for $Y_{i,t}$ with a short panel, the probabilities here need to be replaced by the conditional probabilities given the factors. Hence, $Y_{i,t}$ is identically distributed across time $t$ only if the factors are time-invariant, which is unlikely in practice. $\blacksquare$
\end{example}

\subsection{Monte Carlo Simulations}
\label{sec: monte carlos}

We now turn to Monte Carlo simulations to evaluate the finite sample performance of our procedure. Our set-up follows Example \ref{exmpl: synthetic control} and the literature on synthetic control with individual data, which are of significant applied interest (see \cite{Abadie/LHour:21:JASA}). This will also be the focus of our empirical application below.

We consider a treated group (0) and $K$ untreated groups, where $K \in \{3,5,7\}$. Each group $j$ has a sample size $n_j \in \{100, 200, 1000\}$. The outcomes for individual $i$ in group $j$, $Y_{ij,t}$, are realized over $T$ periods, but only the first $T_0 = 10$ periods are used to match the treated and untreated groups. 

We generate the pre-treatment outcomes for the untreated groups $j=1,...,K$ using the linear model:
\begin{eqnarray*}
Y_{ij,t} = \mu_{j,t} + \varepsilon_{ij,t},
\end{eqnarray*}
\noindent where $\mu_{j,t} = 0.5 + 0.5(-1)^{j-1}\frac{t}{T} + 0.5 \eta_{j,t}$, $\eta_{j,t} \sim_{i.i.d.} N(0,1)$ and $\varepsilon_{ij,t} \sim_{i.i.d.} N(0,1)$. Thus, the mean outcome for $j$, $\mu_{j,t}$ is composed by a deterministic component and a random component. The former grows over time for odd-numbered $j$, while it decreases over time for even-numbered groups. The random component is fixed throughout simulations. Individual outcomes in a group are noisy observations of the associated means. 

We draw data for the treated group as:
\begin{eqnarray*}
Y_{i0,t} = \mu_{0,t} + \varepsilon_{i0,t},
\end{eqnarray*}
where $\mu_{0,t} = w_{0}' \mu_{t}$, $\mu_t = (\mu_{1,t},...,\mu_{K,t})'$, $w_{0} \in \Delta_{K-1}$, and $\varepsilon_{i0,t} \sim_{iid} N(0,1)$ across $i$ and $t$. Thus, the expected outcomes for the treated group are a weighted (synthetic) combination of expected outcomes from the untreated groups with weights, $w_0$.\footnote{While this may resemble the assumption in the synthetic control literature that the outcomes of the treated group have a perfect match in the untreated ones, there is a major distinction: we impose the matching on the population quantities rather than on their sample counterparts. It is important for us to distinguish between population and sample. Otherwise, there is no inference to be made on $w_0$. For the same reason, the observed outcomes are noisy observations of the population means in our setting.} We consider the setting where $T > K$ and there is no linear dependency between the population means $\mu_{j,t}$, $j=0,...,K$. Hence, the weights $w_0$ are point-identified.\footnote{As for our grid for $w$, we use two complementary approaches that allow us to extensively explore both the interior and the boundary of the simplex. The first draws a fine grid of $w$ uniformly over its simplex using a procedure based on \cite{Rubin:81:AoS}. This is done by first drawing a vector of dimension $K-1$, with each element i.i.d. from the uniform distribution with support $[0,1]$. Then, we include 0 and 1 into that drawn vector, sort it, and produce the vector of differences across adjacent elements of $w$. These are all nonnegative and sum up to 1 by construction. As for the boundary points, we first define a step size, generate a vector of $[0,1]$ with points separated by that step-size, and then generate a $K$-dimensional meshed grid of all possible combinations of those values. We only keep the points that (i) add up to 1, (ii) have a zero element. The final grid is the union of the gridpoints generated by both approaches.} 

The weight $w_0$ is the key object in inference. We consider two main specifications, each one presenting a different case of whether $w_0$ is interior or on the boundary of the simplex. In the former (Specification 1), we set $w_0 = (0.2, 0.8 \mathbf{1}_{K-1}'/(K-1))'$, so that $w_0$ lies in the interior of $\Delta_{K-1}$, while, in the latter (Specification 2), $w_0 = (0.5, 0.5, \mathbf{0}_{K-2}')'$, so that $w_0$ lies on the boundary of $\Delta_{K-1}$, where $\mathbf{1}_{d}$ and $\mathbf{0}_{d}$ are $d$-dimensional vectors of 1's and 0's, respectively. In each of $1,000$ simulations, we independently draw $(\varepsilon_{i0,t}, \varepsilon_{i1,t},...,\varepsilon_{K,t})'$ for $t=1,...,T_0$ and $i=1,...,n_j$, while keeping $\mu_{j,t}$ fixed. Since the outcomes $Y_{ij,t}$'s are independent across time, the data generating process represents a repeated cross-section design. All results are shown in Table \ref{mc1}. 


\begin{table}[t]
\caption{\small The Empirical Coverage Probability for $w_0$ at 95\% Nominal Level}
\label{mc1}

\begin{centering}
	\small
	\begin{tabular}{l | ccc | ccc}
    \hline \hline 

 & \multicolumn{3}{c}{Interior $w_0$} & \multicolumn{3}{c}{Boundary $w_0$} 
\\
		\cline{2-4} \cline{5-7}	
		
	  & $K=3$ & $K=5$ & $K=7$ & $K=3$ & $K=5$ & $K=7$
	\\
		\hline 		
    \\

	$n_j=100$ &   0.962   & 0.936  &  0.939  &  0.968  &  0.941 &   0.946
	\\
	$n_j=200$ & 0.947 &   0.946  &  0.941 &   0.949  &  0.946 &  0.939
	\\
	$n_j=1000$ & 0.957  &  0.950  &  0.958  &  0.954  &  0.946 &   0.944  \tabularnewline
	\\
\hline
	\end{tabular}
	\par\end{centering}
    \bigskip
\parbox{6.2in}{\footnotesize
	Notes: The table reports the empirical coverage probability of the confidence set for $w_0$ from our procedure at level $1-\alpha = 0.95$ across two main specifications: the first with an interior $w_0$ and the second with a $w_0$ on the boundary of the simplex. Details are provided in the main text. The number of simulations is $1,000$.} 
\end{table}

Table \ref{mc1} shows that our procedure works very well in finite samples across scenarios. Let us start with the interior weights case (Specification 1). In line with the theory,  coverage probabilities are very close to nominal levels across specifications. This holds even for the smallest sample sizes ($n_j=100, K = 3$ and $n_j = 100, K = 5$), with empirical coverage probabilities between 93.6-96.2\%. Coverage probabilities become very close to 95\% even when $n_j = 200$, across specifications (94.7, 94.6 and 94.1\% for $K \in \{3,5,7\}$, respectively). Similarly, the results continue to hold as $n_j$ or $K$ grows because our asymptotic results are based on the total sample size (numbers of individuals times number of groups). This is seen both with $K=5$ and $K=7$, with coverage probabilities at or close to 95\% for $n_j = 1,000$.

The results from Specification 2 show that our procedure also works well even when $w_0$ is on the boundary of the simplex. Indeed, coverage probabilities are close to the nominal 95\% across specifications and sample sizes. This holds both for small sample sizes (96.8\% and 94.1\% for the smallest values of $K$ and $n_j$), but also as both grow. Indeed, they become very close to 95\% for large sample sizes: with $n_j = 200$, coverage probabilities are 94.9\% for $K=3$, 94.6\% when $K=5$ and 93.9\% for $K=7$. Similar results hold when $n_j = 1,000$.
 
Sometimes researchers are interested in inference on an individual component of $w_0$ rather than inference on the full vector $w_0$. For example, one may be interested in the weight that the synthetic method gives to a specific untreated group (e.g., state, country, firm, or forecaster in the case of predictions). To that end, one can combine our approach with projection-based subvector inference. Table \ref{mc_proj} presents the results of this approach for the first dimension of $w_0$ in the specifications of Table \ref{mc1}. We report both the empirical coverage probability for $w_{0,1}$ and the mean length of the confidence interval. This also provides information on the power of our inference approach.

\begin{table}[t]
\caption{\small Projection-Based Inference for the First Dimension of $w_0$: Empirical Coverage Probability and Length of the Confidence Interval for Scenario 1 Specifications}
\label{mc_proj}
\begin{centering}
	\small
	\begin{tabular}{l|ccc|ccc}
		\hline \hline
        
	&	\multicolumn{3}{c}{Interior $w_0$} &	\multicolumn{3}{c}{Boundary $w_0$}
	\\
		\cline{2-7}

	 & $K=3$ & $K=5$ & $K=7$ & $K=3$ & $K=5$ & $K=7$  \\  
\hline
		\\
		& \multicolumn{6}{c}{Coverage Probability} 	 \\ 
\cline{3-6}
\\
	$n_j=100$ &     0.985 &    0.990  &  0.996   &  0.983 &    0.963 &    0.948
	\\
	$n_j=200$ &     0.969 &    0.984 &    0.994 &    0.972 &   0.953 &    0.940
	\\
	$n_j=1000$ &     0.974  &  0.955  &  0.964  &  0.965  &  0.946  &  0.944 \tabularnewline
	\\
 &\multicolumn{6}{c}{Length of the Confidence Interval} \\
\cline{3-6}
\\
	$n_j=100$ & 0.163&    0.239 &   0.340 &   0.136 &   0.067 &   0.114
	\\
	$n_j=200$ &     0.112  &  0.147 &   0.236 &   0.094 &   0.030 &   0.023
	\\
	$n_j=1000$ &     0.037 &   0.012 &  0.020  &  0.029     &    0.000   &      0.000 \tabularnewline
    \\
	\hline
	\end{tabular}
	\par\end{centering}
    \bigskip
\parbox{6.2in}{\footnotesize
	Notes: The top panel reports the empirical coverage probability of the confidence interval for the projection-based inference for the first dimension of $w_0$ in the specifications of Table \ref{mc1} for $1-\alpha = 0.95$. Details are provided in the main text. The bottom panel reports the associated mean length of the confidence intervals across $R=1,000$ simulations. The latter is computed only on non-empty confidence intervals, thereby making the results conservative.} 
\end{table}

The projection-based inference works well in finite samples. As expected, it is more conservative than inference on the full $w_0$ (c.f. Table \ref{mc1}). For $K=3, 5$ and interior $w_0$, the former is typically 2-4 percentage points more conservative than the latter, although such conservativeness decreases with sample size (e.g., it is below 1 percentage point when $n_j = 1,000$ and $K=5$ or $K=7$, regardless of specification). In spite of its conservative nature, projection-based inference is still informative in our setting. Even when sample size is small ($n_j  = 100, K=3$), the average length of the confidence interval can be as short as 0.14-0.16 and it is far from spanning $[0,1]$. Furthermore, for a fixed $K$ and weight $w_0$, as sample size $n$ increases, the length of the confidence interval quickly decreases and gets even close to 0: when $n_j = 1,000$, the average length is below 0.04 across specifications. The latter would be very sharp for inference with a sample size often found in empirical applications with individual-level data. Thus, our simulations suggest that, when $w_0$ is point-identified and the sample size is large enough, projection-based inference for the weights can be informative and not overly conservative. 

\section{Empirical Application}
\label{sec: empirical app}

As an empirical illustration of our method, we study how a large increase in minimum wages in Alaska in 2003, from US\$5.65 to US\$7.15, affected average family income. To do so, we use individual-level data and state-level weights, as in Example \ref{exmpl: synthetic control}. This follows a recent strand of the literature that uses synthetic control methods to obtain credible estimates of the effects of minimum wage increases on economic outcomes (e.g., \cite{Allegretto/Dube/Reich/Zipperer:17:ILR}, \cite{Neumark/Wascher:17:ILR}, \cite{Powell:22:JBES}) and the analysis of this policy in \cite{Gunsilius:23:ECMA}. 

Following \cite{Dube:19:AEJ} and \cite{Gunsilius:23:ECMA}, our outcome of interest is a measure of household income, called average family income. This is a measure which adjusts for family size and composition (i.e., ``equivalized'') and it is measured in multiples of the federal poverty threshold. We use the main sample from \cite{Dube:19:AEJ} (i.e., individuals aged under 65), originally drawn from the Current Population Survey (CPS), and focus on the subsample from 1998-2002 (i.e., the five years before the policy). 

To form a synthetic control, we form a pool of states that resemble Alaska in terms of population and economic activity. We start with states that, like Alaska, are among the top-10 oil producing states in 2002 (i.e., prior to the policy), e.g., see Figure 7 in \cite{Ismayilova:07:Thesis}. We then drop states that have a minimum wage increase during the period of interest (California) and states that have a population over 3 million (Texas, Louisiana and Oklahoma). By comparison, Alaska had a population of approximately 650,000 in 2002, \cite{Census2002}. Our final set of control states is, thus: Kansas, Mississippi, New Mexico, North Dakota and Wyoming ($K=5$). From this pool, we construct a synthetic comparison state by deriving state-level weights from estimating $\mu_{j,t}$: the mean family income in each state $j$ and year $t$.

We use our statistical procedure outlined in Section \ref{subsec: simple confidence set} and report the projection-based inference for weights and Bonferroni-based confidence interval for a parameter of interest, $\theta_0$, defined as the difference in mean outcomes in Alaska in 2003 (the year post-policy) relative to the mean in the synthetic control. This can be written as:
\begin{eqnarray}
\theta(w)= \mu_{\text{treat},2003} - \mu_{\text{ctrl},2003}' w,\label{emp_bonf}
\end{eqnarray}
where $\mu_{\text{ctrl},2003} = [\mu_{1,2003}, \mu_{2,2003},...,\mu_{K,2003}]'$ is a vector of the mean average family income across families in the control states and $\theta_0 = \theta(w_0)$. We can compute the asymptotic variance $\Omega(w)$ as explained in Example \ref{exmpl: synthetic control30}.

\begin{table}[t!]
\caption{\small The Effects of Alaska's Minimum Wage Increase on Average Family Income}
\label{emp1}
\begin{centering}
	\small
	\begin{tabular}{lcccc}
		\hline \hline
	 & \multicolumn{2}{c}{Specification 1} & \multicolumn{2}{c}{Specification 2} \\
\\
   Inference for $\theta_0$:     & Bonferroni & Projection  & Bonferroni & Projection \\
     
\hline
		\\
\\
	$\theta_0$ - Change in  & $[-0.462, 0.015]$ &  $[-0.392, 0.014]$  & $[-0.462, 0.031]$ & $[-0.392, 0.010]$ \tabularnewline
Average Family Income	\\
\\
	$w_0$ - Kansas  & $[0.825, 1]$  & $[0.625, 0.950]$ & $[0.750, 1]$ &$ [0.650, 1]$\\
		\\
	$w_0$ - Mississippi  & -  & - & $[0, 0.100]$ & $[0, 0]$ \\
\\
	$w_0$ - New Mexico & $[0, 0.100]$ & $[0, 0]$ & $[0, 0.100]$ &$ [0, 0]$\\
\\
	$w_0$ - North Dakota & $[0, 0.125]$ & $[0, 0]$ & $[0, 0.150]$ & $[0, 0]$\\
\\
	$w_0$ - Wyoming & $[0, 0.175]$ & $[0.050, 0.375]$ & $[0, 0.250]$  & $[0.050, 0.350]$\\
	\\
	\hline
	\end{tabular}
    \bigskip
	\par\end{centering}
\parbox{6.2in}{\footnotesize
	Notes: The table reports the results of our empirical application on the effects of the minimum wage increase in Alaska in 2003 on average family income (measured in multiples of the federal poverty threshold and adjusted for family size and composition). The mean outcome for the treated state (Alaska) is 3.427 (pre-2003) and 3.309 (in 2003). The first row presents the confidence interval for the effect in 2003, after the policy, with $\alpha = 0.05$, while the other rows report the projection-based inference confidence interval at level $1-\alpha$ for the weights. We present the results for two specifications, using both the Bonferroni method ($\kappa = 0.005$) from Example \ref{exmpl: synthetic control30} and the projection method described in Example \ref{exmpl: synthetic control30} and obtained from (\ref{CI theta2}). Specification 2 uses the full set of control states, while Specification 1 drops Mississippi. } 
	\end{table}

	The first row of Table \ref{emp1} presents the confidence interval for the effect of the increase in minimum wages on average family income in 2003 across specifications. The results are robust across specifications and inference methods, with confidence intervals centered around -0.2. From our results, we can reject the presence of any meaningful positive effects, as the upper bound of our confidence intervals are all very close or at 0 (and below a 1\% effect). This suggests that this minimum wage policy, at best, did not affect average family income. At worst, it decreased average family income by up to 13.4\% relative to a pre-policy (1998-2002) average income of 3.43 times the federal poverty threshold. While it is likely that the poorest households benefit from this policy, the average family may have lost a small amount of income, possibly due to increased unemployment. Our confidence intervals account for inference on the weights, which contrasts with the standard approach in synthetic control, where the synthetic weights are assumed to hold exactly for any realization of the sample. While the results are robust across specifications (i.e., with or without Mississippi) and methods (Bonferroni or projection-based), we do find that the projection method presented in Algorithm \ref{alg:confidence_set3} seems to outperform the Bonferroni counterpart in this setting. Indeed, the average length of the projection-based confidence interval for $\theta_0$ is 16\% (0.4/0.48) shorter than the Bonferroni counterpart, and the former is a subset of the latter. This empirical gain comes at an increased computational cost. 

	Table \ref{emp1} further reports projection-based confidence intervals for weights across two choices of control units: with and without Mississippi. Kansas often received the largest weights (i.e., with confidence intervals centered around 0.8-0.85), followed by Wyoming, while the others have relatively similar confidence intervals centered around 0-0.05. Meanwhile, the average length of the confidence intervals is around 0.15. This provides further support to the findings in the simulations that projection-based inference for $w$ can be informative in this context.

\section{Conclusion}

In this paper, we propose a new method for inference on the simplex-constrained weights. Our method is based on the projection of an asymptotically normal statistic onto a polyhedral cone. We show that the critical value for the test statistic can be read from the $\chi^2$ distribution with data-dependent degrees of freedom. This leads to an asymptotically uniformly valid confidence set for the weights. We provide a proof of this result and show that it is applicable to a wide range of settings. We also show that the method works well in finite samples through Monte Carlo simulations. We apply our method to a synthetic control example and show that it can be useful in practice.

\bibliographystyle{econometrica}
\bibliography{simplex}

   \appendix
   \section*{Appendix: Mathematical Proofs}
   \setcounter{section}{1}
   \setcounter{equation}{0}
   \counterwithin{equation}{section}

Given a symmetric positive definite matrix $\Omega$, we define the norm $\| \cdot \|_\Omega$ as $\|x\|_\Omega = \sqrt{x'\Omega^{-1}x}$ and inner product $\langle x,y \rangle_{\Omega} = x' \Omega^{-1} y$. Given a nonempty closed convex set $C$, take the projection $\Pi_\Omega(y\mid C)$ (along the norm $\| \cdot \|_\Omega$) to be the solution to the following minimization problem:
\begin{align}
	\inf_{x \in C} \|y - x\|_{\Omega}^2.
\end{align}
Since $\Omega$ is positive definite and $C$ is closed and convex, the projection $\Pi_\Omega(y\mid C)$ exists and is unique. For any vector $x \in \mathbf{R}^k$ and a matrix $A \in \mathbf{R}^{k \times k}$, we denote $[x]_j$ to mean its $j$-th entry and $[A]_j$ to mean its $j$-th row vector. For any nonempty $J \subset \{1,...,k\}$, $[x]_J$ denotes the subvector of $x$ whose entries consist of $[x]_j$, $j \in J$, and $[A]_J$ denotes the submatrix of $A$ whose rows consist of $[A]_j$, $j \in J$. For any matrix $A$, $\text{rk}(A)$ denotes the rank of $A$. Given a matrix $A$, $\|A\|$ denotes the Frobenius norm of $A$, i.e., $\|A\| = \sqrt{\text{tr}(A'A)}$.

Let $\overline K := \{1,...,K\}$. For any vector $x \in \mathbf{R}^K$, recall the definition: $J_0[x] := \{j \in \overline K: [x]_j = 0\}.$ The set, $J_0[x]$, consists of the indices of the first $K$ entries of $x$ that are zeros. We assume that any inequality or equality that involves a vector $[x]_J$ with $J = \varnothing$ is vacuously true. Given a symmetric positive definite $(K+L-1) \times (K+L-1)$ matrix, $\Omega$, let 
\begin{align*}
	M := B \Omega^{-1},
\end{align*}
where $B$ is defined in (\ref{B}). Thus, $M$ is a $K \times (K+L-1)$ matrix. Recall the definition of $\tilde \Lambda(w)$ in (\ref{tilde Lambda(w)}). Its polar cone is given by
\begin{align*}
	\tilde \Lambda^\circ(w,\Omega) &:= \left\{x \in \mathbf{R}^{K + L -1} : [Mx]_{J_0[w]} \le 0 \right\}.
\end{align*}
(An inequality between vectors is understood as holding element-wise.) For each $J \subset \overline K$, define 
\begin{align*}
	\tilde \Lambda_J^\circ(w,\Omega) := \left\{x \in \mathbf{R}^{K + L -1} : [M x]_{J} = 0, \enspace [M x]_{J_0[w] \setminus J} \le 0 \right\}.
\end{align*}
Each face of $\tilde \Lambda^\circ(w,\Omega)$ is identical to $\tilde \Lambda_J^\circ(w,\Omega)$ for some $J \subset J_0[w]$. The linear span of $\tilde \Lambda_J^\circ(w,\Omega)$ and its relative interior (relative to its linear span) are given by  
\begin{align}
	\label{ri Lambda J}
	L_J(\Omega) &:= \left\{x \in \mathbf{R}^{K + L -1} : [M x]_{J} = 0 \right\} \text{ and }\\ \notag
	\text{ri}(\tilde \Lambda_J^\circ(w,\Omega)) &:= \left\{x \in \mathbf{R}^{K + L -1}: [M x]_J = 0, \enspace [M x]_{J_0[w] \setminus J} < 0 \right\}.
\end{align}
It is not hard to see that the sets $\text{ri}(\tilde \Lambda_J^\circ(w,\Omega))$, $J \subset J_0[w]$, form a partition of $\tilde \Lambda^\circ(w,\Omega)$. 

\subsection{Preliminary Results}

\begin{lemma}
	\label{lemm: partition}
  For any $w \in \Delta_{K-1}$, $J \subset J_0[w]$, and $x \in \tilde \Lambda^\circ(w,\Omega)$,
  \begin{align*}
    x \in \text{ri}(\tilde \Lambda_J^\circ(w,\Omega)) \text{ if and only if } J_0[Mx] \cap J_0[w] = J.
  \end{align*}
\end{lemma}

\begin{proof} The result follows because the sets, $\text{ri}(\tilde \Lambda_J^\circ(w,\Omega))$, $J \subset J_0[w]$, partition $\tilde \Lambda^\circ(w,\Omega)$.
\end{proof}

\begin{lemma}
    \label{lemm: change of var} 
    For any vector $y \in \mathbf{R}^{K+L-1}$ and $w \in \Delta_{K-1}$,
   \begin{align*}
    \Pi_{\Omega}(y \mid \tilde \Lambda^\circ(w,\Omega)) = \Omega \Pi_{\Omega^{-1}}(\Omega^{-1} y \mid \tilde \Lambda^\circ(w,I)).
   \end{align*}
\end{lemma}

\begin{proof}
   Note that 
   \begin{align*}
      \Pi_{\Omega}(y \mid \tilde \Lambda^\circ(w,\Omega)) &= \argmin_{x: [M x]_{J_0[w]} \le 0} (y - x)'\Omega^{-1} (y - x)\\
      &= \Omega \argmin_{x: [M \Omega x]_{J_0[w]} \le 0} (\Omega^{-1} y - x)' \Omega (\Omega^{-1}y - x)\\
      &= \Omega \argmin_{x: [B x]_{J_0[w]} \le 0} (\Omega^{-1} y - x)' \Omega (\Omega^{-1}y - x) = \Omega\Pi_{\Omega^{-1}}(\Omega^{-1} y \mid \tilde \Lambda^\circ(w,I)).
   \end{align*}
\end{proof}

\begin{lemma}
    \label{lemm: rank}
   For all nonempty $J \subset \overline K$ such that $|J| \le K-1$, 
   \begin{align*}
    \text{rk}([B_2]_J) = |J|.
   \end{align*} 
\end{lemma}

\begin{proof} 
Let $A_J = S_J(I - \mathbf{1}\mathbf{1}'/K)S_J'$, where $S_J$ is a $J \times K$ selection matrix such that $S_J U$ for any matrix $U$ selects the rows of $U$ indexed by $J$. Since $B_2 B_2' = I - \mathbf{1}\mathbf{1}'/K$ and $[B_2]_J = S_J B_2$, $\text{rk}([B_2]_J) = \text{rk}(A_J).$ For each $i \in J$, let $d_i$ denote the column vector of $A_J$ which is a subvector of the $i$-th column vector of $I - \mathbf{1}\mathbf{1}'/K$. Suppose to the contrary that $\text{rk}(A_J) < |J|$. Then there exists $i \in J$ such that, with $J_{-i} := J \setminus \{i\}$,  
\begin{align}
    \label{equations}
    d_i = \sum_{j \in J_{-i}} a_j d_j,
\end{align}
for some constants $a_j$. Let $\tilde e_i$ be the $K$ dimensional vector whose $i$-th entry is one and the other entries zero. For each $i \in J$, let $e_i = S_J \tilde e_i$. Note that $d_i = e_i - \mathbf{1}/K \in \mathbf{R}^{|J|}$. The equality in (\ref{equations}) involves $|J|$ equations, where the $i$-th equation and the $\ell$-th equation with $\ell \in J_{-i}$ take the following form: 
\begin{align*}
    1 - \frac{1}{K} = - \sum_{j \in J_{-i}} \frac{a_j}{K} \text{ and } - \frac{1}{K} = a_\ell - \sum_{j \in J_{-i}} \frac{a_j}{K}.
\end{align*}  
This implies that $a_\ell = -1$ for all $\ell \in J_{-i}$. The first equation above then yields 
\begin{align*}
    1 - \frac{1}{K} = \frac{|J|-1}{K},
\end{align*} 
or $|J| = K$, which contradicts the condition that $|J| \le K-1$.
\end{proof}

\begin{lemma}
    \label{lemm: rank2}
   For each $J \subset \overline K$, let $P_J$ be the $(K+L-1) \times (K+L-1)$ projection matrix onto $L_J(\Omega)$ (along the norm $\|\cdot \|_\Omega$). If $|J| \le K-1$, then $\text{rk}(P_J) = K + L - 1 - |J|.$
\end{lemma}

\begin{proof} 
    If $J = \varnothing$, $L_J(\Omega) = \mathbf{R}^{K + L -1}$ and $\text{rk}(P_J) = K + L -1$. Suppose that $J \ne \varnothing$. Note that 
\begin{align*}
	L_J(\Omega) = \left\{x \in \mathbf{R}^{K + L -1}: [M x]_J = 0 \right\} = \left\{x \in \mathbf{R}^{K + L -1}: [M]_J x = 0 \right\}.
\end{align*} 
Let $(\Omega^{-1})_{11}$ denote the first $(K-1) \times (K-1)$ block diagonal submatrix and $(\Omega^{-1})_{12}$ the $(K-1) \times L$ block submatrix of $\Omega^{-1}$. Since $[(\Omega^{-1})_{11},(\Omega^{-1})_{12}]$ has full row rank, 
\begin{align}
	\label{eq M B2}
	\text{rk}\left([M]_J\right) = \text{rk}\left([ B_2(\Omega^{-1})_{11}, B_2(\Omega^{-1})_{12} ]_J\right) = \text{rk}([B_2]_J),
\end{align}
Hence, 
\begin{align*}
	\text{rk}(P_J) = K + L -1 - \text{rk}([M]_J) = K  + L - 1 - \text{rk}([B_2]_J) = K + L - 1 - |J|,
\end{align*}
by Lemma \ref{lemm: rank}.
\end{proof}

The following lemma gives a characterization of the critical value in Theorem 1 of  \cite{AlMohamad/vanZwet/Cator/Goeman:20:Biometrika}. For notational brevity, we write 
\begin{align*}
    \pi_{\Omega}^\circ(y;w) := M\Pi_{\Omega}(y \mid \tilde \Lambda^\circ(w,\Omega)), \enspace y \in \mathbf{R}^{K+L-1}.
\end{align*}

\begin{lemma}
	\label{lemm: Mohamad}
   Suppose that $Y \in \mathbf{R}^{K+L-1}$ is a random vector following $N(\mu,\Omega)$, with a symmetric positive definite $(K+L-1)\times (K+L-1)$ matrix $\Omega$. Then, for any $\alpha \in (0,1)$ and $w \in \Delta_{K-1}$,
\begin{align}
	\label{eq: rej prob}
	P\left\{\left\|  Y - \Pi_\Omega(Y\mid \tilde \Lambda(w))\right\|_\Omega^2 > c_{1-\alpha,\Omega}(Y;w) \right\} \le \alpha,
\end{align}
where 
\begin{align}
	\label{eq: critical value}
	c_{1-\alpha,\Omega}(y;w) = G^{-1}\left(1-\alpha; \max\left\{K+L-1 - \left|J_0[\pi_\Omega^\circ(y;w)] \cap J_0[w] \right|,1\right\} \right), \enspace y \in \mathbf{R}^{K + L -1},
\end{align} 
and $G(\cdot; k)$ is the CDF of the $\chi^2$ distribution with degrees of freedom equal to $k$.
\end{lemma}

\begin{proof} Let $F_\ell$, $\ell=1,...,L$, be the faces of the polyhedral cone $\tilde \Lambda^\circ(w,\Omega)$, and let $\text{ri}(F_\ell)$ be the relative interior of $F_{\ell}$. Then, by Theorem 1 of  \cite{AlMohamad/vanZwet/Cator/Goeman:20:Biometrika},\footnote{Note that they apply Lemma 3.13.2 of \cite{Silvapulle/Sen:05:ConstrainedStatInference}, p.125, which uses the orthogonal decomposition of $\mathbf{R}^K$ equipped with the inner product $\langle a,b\rangle = a'b$. However, the lemma continues to hold with any other inner product, with the definition of projections and orthogonal complements appropriately redefined.} we have
\begin{align}
	\label{eq: rej prob1}
	P\left\{\left\| Y - \Pi_\Omega(Y\mid \tilde \Lambda(w))\right\|_\Omega^2 > q_{1-\alpha,\Omega}(Y;w)\right\} \le \alpha,
\end{align}
where
\begin{align}
	\label{q}
	q_{1-\alpha,\Omega}(Y;w) = \sum_{\ell =1}^L 1\{\Pi_\Omega(Y\mid \tilde \Lambda^\circ(w,\Omega)) \in \text{ri}(F_\ell)\} G^{-1}\left(1-\alpha; \max\{\text{rk}(P_\ell),1\}\right),
\end{align}
and $P_\ell$ denotes the projection matrix (along $\|\cdot \|_\Omega$) onto the linear span of $F_\ell$. It remains to show that 
\begin{align*}
	q_{1-\alpha,\Omega}(Y;w)  = c_{1-\alpha,\Omega}(Y;w).
\end{align*}

In our case with the polyhedral cone $\tilde \Lambda^\circ(w,\Omega)$, the faces and their relative interiors are given by $\tilde \Lambda_J^\circ(w,\Omega)$ and $\text{ri}(\tilde \Lambda_J^\circ(w,\Omega))$, $J \subset J_0[w]$ (see, e.g., the proof of Lemma 3.13.5 of \cite{Silvapulle/Sen:05:ConstrainedStatInference}). Hence, we can rewrite $q_{1-\alpha,\Omega}(Y;w)$ as
\begin{align}
	\label{sum}
	 &\sum_{J \subset J_0[w]} 1\left\{\Pi_\Omega(Y\mid \tilde \Lambda^\circ(w,\Omega)) \in \text{ri}(\tilde \Lambda_J^\circ(w,\Omega))\right\} G^{-1}\left(1-\alpha; \text{max}\left\{\text{rk}(P_J),1\right\}\right),
\end{align}
where $P_J$ is the projection matrix onto $L_J(\Omega)$. Consider $J \subset J_0[w]$. Since $w \in \Delta_{K-1}$, we have $|J| \le K-1$. By Lemma \ref{lemm: rank2}, $\text{rk}(P_J) = K + L - 1 - |J|$. Since the relative interiors partition $\tilde \Lambda^\circ(w,\Omega)$, there exists a unique $J^*(Y) \subset J_0[w]$ such that 
\begin{align*}
    \Pi_\Omega(Y\mid \tilde \Lambda^\circ(w,\Omega)) \in \text{ri}\left(\tilde \Lambda_{J^*(Y)}^\circ(w,\Omega)\right)
\end{align*}
 and, by Lemma \ref{lemm: partition}, $J^*(Y) = J_0[\pi_{\Omega}^\circ(Y;w)] \cap J_0[w]$. Hence, from (\ref{sum}),
\begin{align*}
	q_{1-\alpha,\Omega}(Y;w) = G^{-1}\left(1-\alpha; \max\left\{K + L -1 - |J_0[\pi_{\Omega}^\circ(Y;w)] \cap J_0[w]|,1\right\} \right) = c_{1-\alpha,\Omega}(Y;w).
\end{align*}
\end{proof}

\begin{lemma}
	\label{lemm: conv0}
	Suppose that $x_n$ and $\tilde x_n$ are sequences of vectors in $\mathbf{R}^{K + L -1}$, and $\Omega_n$ and $\tilde \Omega_n$ are sequences of $(K + L -1) \times (K + L -1)$ symmetric positive definite matrices such that $\| x_n - \tilde x_n \| \rightarrow 0 \text{ and } \| \Omega_n - \tilde \Omega_n \| \rightarrow 0$, as $n \rightarrow \infty$, where
	\begin{align}
		\label{min eigenvalue}
		\limsup_{n \rightarrow \infty} \|\Omega_n\| < \infty \text{ and } \liminf_{n \rightarrow \infty} \lambda_{\min}(\Omega_n) >0.
	\end{align} 
    Let $C_n \subset \mathbf{R}^{K + L -1}$ be a sequence of nonempty closed convex cones. Then,  
    \begin{align*}
        \left\|\Pi_{\Omega_n}(x_{n} \mid C_n) - \Pi_{\Omega_n}(\tilde x_n \mid C_n)\right\| \rightarrow 0, \text{ as } n \to \infty.
    \end{align*}
    
    Suppose further that either 
    
    \quad \quad (a) $\limsup_{n \rightarrow \infty} \| \Pi_{\Omega_n}(x_{n} \mid C_n) \| < \infty$, or 
    
    \quad \quad (b) $\limsup_{n \rightarrow \infty} \| x_n - \Pi_{\Omega_n}(x_{n} \mid C_n) \| < \infty$. 
    
    Then,
    \begin{align}
        \label{conv0}
        \left\|\Pi_{\Omega_n}(x_{n} \mid C_n) - \Pi_{\tilde \Omega_n}(\tilde x_n \mid C_n)\right\| \rightarrow 0, \text{ as } n \rightarrow \infty.
    \end{align}
\end{lemma}

\begin{proof} 
Let us prove the first statement. Since a projection map on a closed convex set is a contraction map (see, e.g. Theorem 3 of \cite{Cheney/Goldstein:59:PAMS}), 
\begin{align*}
	\left\|\Pi_{\Omega_{n}}(x_{n} \mid C_n) - \Pi_{\Omega_{n}}(\tilde x_n \mid C_n)\right\|_{\Omega_n} \le \| x_{n} - \tilde x_n \|_{\Omega_{n}} \rightarrow 0,
\end{align*}
as $n \rightarrow \infty$, by the conditions of the lemma and (\ref{min eigenvalue}).

Let us turn to proving the second statement. Note that
\begin{align}
	\label{conv31}
	\left\|\Pi_{\Omega_{n}}(x_{n} \mid C_n) - \Pi_{\tilde \Omega_n}(\tilde x_n \mid C_n)\right\| \le A_{n,1} + A_{n,2},
\end{align}
where  
\begin{align*}
	A_{n,1} = \left\|\Pi_{\Omega_{n}}(x_{n} \mid C_n) - \Pi_{\Omega_{n}}(\tilde x_n \mid C_n)\right\| \text{ and }
	A_{n,2} = \left\|\Pi_{\Omega_{n}}(\tilde x_n \mid C_n) - \Pi_{\tilde \Omega_n}(\tilde x_n \mid C_n)\right\|.
\end{align*}
By the first statement of the lemma, we have $\lim_{n \rightarrow \infty} A_{n,1} = 0$. We deal with $A_{n,2}$. Let $C_n^\circ$ denote the polar cone of $C_n$. Without loss of generality, we assume that Condition (a) holds. (The case with Condition (b) can be dealt with by interchanging $C_n^\circ$ with $C_n$ in the proof below.) For simplicity, let 
\begin{align*}
    a_{n} = \Pi_{\Omega_{n}}(\tilde x_{n} \mid C_{n}^\circ) \text{ and } \tilde a_{n} = \Pi_{\tilde \Omega_n}(\tilde x_{n} \mid C_{n}^\circ).
\end{align*} 
Since $\tilde x_n = \Pi_{\Omega_{n}}(\tilde x_{n} \mid C_{n}) + \Pi_{\Omega_{n}}(\tilde x_{n} \mid C_{n}^\circ)$, it suffices to show that 
\begin{align*}
   A_{n,2} = \| a_n - \tilde a_n \| \rightarrow 0, \text{ as } n \to \infty.
\end{align*}

Suppose to the contrary that there exists a subsequence $\{n'\}$ such that $\lim_{n' \rightarrow \infty} A_{n',2} >0$. Let $\epsilon>0$ be such that $0 < \epsilon < \lim_{n' \rightarrow \infty} A_{n',2}.$ Let
\begin{align}
    \label{delta}
    c_{n'} := \min \left\{\lambda_{\min}(\Omega_{n'}^{-1}),\lambda_{\min}(\tilde \Omega_{n'}^{-1})\right\}.
\end{align}

Now, since $\|a_{n'} - \tilde a_{n'}\|_{\tilde \Omega_{n'}}^2 \ge c_{n'} \|a_{n'} - \tilde a_{n'}\|^2 > c_{n'} \epsilon^2$, note that 
\begin{align}
    \label{ineq312}
    \|\tilde x_{n'} - a_{n'}\|_{\tilde \Omega_{n'}}^2 &= \|\tilde x_{n'} - \tilde a_{n'}\|_{\tilde \Omega_{n'}}^2 + \|a_{n'} - \tilde a_{n'}\|_{\tilde \Omega_{n'}}^2 + 2\langle \tilde a_{n'} - \tilde x_{n'},a_{n'} - \tilde a_{n'} \rangle_{\tilde \Omega_{n'}}\\ \notag
    &\ge \|\tilde x_{n'} - \tilde a_{n'}\|_{\tilde \Omega_{n'}}^2 + c_{n'}\epsilon^2,
\end{align}
by the variational inequality in the convex optimization. Similarly,
\begin{align}
    \label{ineq313}
    \|\tilde x_{n'} - \tilde a_{n'}\|_{\Omega_{n'}}^2 \ge \|\tilde x_{n'} - a_{n'}\|_{\Omega_{n'}}^2 + c_{n'}\epsilon^2.
\end{align}
Note that for any vector $a \in \mathbf{R}^{K+L-1}$, 
\begin{align*}
	\|a\|_{\tilde \Omega_{n'}}^2 - \|a\|_{\Omega_{n'}}^2 &= a'(\tilde \Omega_{n'}^{-1} - \Omega_{n'}^{-1}) a\\
                                                         &= a'\Omega_{n'}^{-1/2}(\Omega_{n'}^{1/2} \tilde \Omega_{n'}^{-1} \Omega_{n'}^{1/2} - I)\Omega_{n'}^{-1/2} a\\
														 &= \text{tr}\left( (\Omega_{n'}^{1/2} \tilde \Omega_{n'}^{-1} \Omega_{n'}^{1/2} - I)\Omega_{n'}^{-1/2} a a'\Omega_{n'}^{-1/2} \right)\\
														 &\ge \lambda_{\min}\left(\Omega_{n'}^{1/2} \tilde \Omega_{n'}^{-1} \Omega_{n'}^{1/2} - I\right) \|a\|_{\Omega_{n'}}^2,
\end{align*}
Define $\tilde \delta_{n'} = \lambda_{\min}\left(\Omega_{n'}^{1/2} \tilde \Omega_{n'}^{-1} \Omega_{n'}^{1/2} - I\right)$ and $\delta_{n'} = \lambda_{\min}\left(\tilde \Omega_{n'}^{1/2} \Omega_{n'}^{-1} \tilde \Omega_{n'}^{1/2} - I\right)$. Then, for any vector $a \in \mathbf{R}^{K + L -1}$, we have 
\begin{align}
	\label{ineqs354}
	\|a\|_{\tilde \Omega_{n'}}^2 \ge ( 1 + \tilde \delta_{n'}) \| a\|_{\Omega_{n'}}^2 \text{ and } \|a\|_{\Omega_{n'}}^2 \ge ( 1 + \delta_{n'}) \| a\|_{\tilde \Omega_{n'}}^2. 
\end{align}
Now, observe that
\begin{align}
    \|\tilde x_{n'} - a_{n'}\|_{\tilde \Omega_{n'}}^2 - c_{n'}\epsilon^2 &\ge \|\tilde x_{n'} - \tilde a_{n'}\|_{\tilde \Omega_{n'}}^2 \enspace (\text{by (\ref{ineq312})}) \\ \notag
	&\ge (1 + \tilde \delta_{n'}) \|\tilde x_{n'} - \tilde a_{n'}\|_{\Omega_{n'}}^2 \enspace (\text{by (\ref{ineqs354})})\\ \notag
	&\ge (1 + \tilde \delta_{n'}) (\|\tilde x_{n'} - a_{n'}\|_{\Omega_{n'}}^2  + c_{n'} \epsilon^2) \enspace (\text{by (\ref{ineq313})})\\ \notag
	&\ge (1 + \tilde \delta_{n'}) ((1 + \delta_{n'}) \|\tilde x_{n'} - a_{n'}\|_{\tilde \Omega_{n'}}^2  + c_{n'} \epsilon^2).
\end{align} 
Hence, \begin{align*}
	\|\tilde x_{n'} - a_{n'}\|_{\tilde \Omega_{n'}}^2 \ge  (1 + \tilde \delta_{n'})(1 + \delta_{n'}) \|\tilde x_{n'} - a_{n'}\|_{\tilde \Omega_{n'}}^2  + (1 + \tilde \delta_{n'})c_{n'} \epsilon^2 + c_{n'}\epsilon^2.
\end{align*}
Using this, we deduce that 
\begin{align}
	\label{der3}
	\|\tilde x_{n'} - a_{n'}\|_{\Omega_{n'}}^2 &\ge (1 + \delta_{n'})\|\tilde x_{n'} - a_{n'}\|_{\tilde \Omega_{n'}}^2 \enspace (\text{by (\ref{ineqs354})})\\ \notag
                     &\ge (1 + \delta_{n'})^2(1 + \tilde \delta_{n'}) \|\tilde x_{n'} - a_{n'}\|_{\tilde \Omega_{n'}}^2  + (1 + \delta_{n'})(1 + \tilde \delta_{n'})c_{n'} \epsilon^2\\ \notag
					 &\quad \quad + (1 + \delta_{n'})c_{n'}\epsilon^2\\ \notag
					 &\ge (1 + \delta_{n'})^2 (1 + \tilde \delta_{n'})^2  \|\tilde x_{n'} - a_{n'}\|_{\Omega_{n'}}^2  + (1 + \delta_{n'}) (1 + \tilde \delta_{n'}) c_{n'} \epsilon^2 \\ \notag
					 &\quad \quad + (1 + \delta_{n'}) c_{n'}\epsilon^2 \enspace (\text{by (\ref{ineqs354})}).
\end{align}
Since $\lim_{n' \rightarrow 0} A_{n',1} = 0$, we have 
\begin{align*}
	\limsup_{n' \rightarrow \infty}\|\tilde x_{n'} - a_{n'}\| &= \limsup_{n' \rightarrow \infty}\|\Pi_{\Omega_{n'}}(\tilde x_{n'} \mid C_{n'})\|\\
	 &=\limsup_{n' \rightarrow \infty}\|\Pi_{\Omega_{n'}}(x_{n'} \mid C_{n'})\| < \infty,
\end{align*}
by Condition (a). By (\ref{min eigenvalue}) and the condition $\|\Omega_n - \tilde \Omega_n\| \rightarrow 0$ as $n \rightarrow \infty$, we have $\epsilon' : = \liminf_{n \rightarrow \infty} c_n >0$. As $n' \rightarrow \infty$, we have $\delta_{n'}, \tilde \delta_{n'} \rightarrow 0$. Hence, 
\begin{align*}
	\delta_{n'}\|\tilde x_{n'} - a_{n'}\|_{\Omega_{n'}}^2 \le \delta_{n'} \lambda_{\max}(\Omega_{n'}^{-1}) \|\tilde x_{n'} - a_{n'}\|^2 \rightarrow 0,
\end{align*} 
as $n' \rightarrow 0$. Similarly, $\tilde \delta_{n'}\|\tilde x_{n'} - a_{n'}\|_{\Omega_{n'}}^2 \rightarrow 0$. From (\ref{der3}), this yields $2 \epsilon' \epsilon^2 \le 0$, a contradiction. Hence, $\lim_{n \rightarrow \infty} A_{n,2} = 0$. 
\end{proof}

\subsection{Proofs of the Main Results}

We first prove Lemma \ref{lemm: Mohamad2}. For each $J \subset \overline K$, let $\tilde P_{n,J}$ denote the projection matrix onto $L_J(I)$ along $\langle \cdot,\cdot \rangle_{\tilde \Omega_{n}^{-1}}$ and $\tilde P_{J}$ the projection matrix onto $L_J(I)$ along $\langle \cdot,\cdot \rangle_{\Omega^{-1}}$. Due to the almost sure representation theorem (cf. Theorem 6.7 of \cite{Billingsley:99:ConvMeas}, p.70), there is a common probability space, say, $(\Upsilon,\mathcal{F},\mathbf{P})$, on which we have a sequence of random vectors $\tilde Z_n$ and random matrices $\tilde \Omega_n$ such that
\begin{align}
    \label{almost sure}
	[\tilde Z_n^{\prime},\text{vec}'(\tilde \Omega_n - \Omega_n)]^{\prime} \rightarrow_{a.s} [\tilde Z^{\prime}, 0']^{\prime},
\end{align}
as $n \rightarrow \infty$, where $[\tilde Z_n', \text{vec}'(\tilde \Omega_n)]'$ and $\tilde Z$ has the same distribution as $[Z_n', \text{vec}'(\hat \Omega_n)]'$ and $N(0,I)$ respectively. Note that $\Omega_n$ is bounded, with its minimum eigenvalue bounded away from zero. Furthermore, $J_0[w_n] \subset \overline K$ for all $n \ge 1$. Hence, for any subsequence of $\{n\}$, there exists a further subsequence $\{n'\}$ and $\overline J,\overline J' \subset \overline K$ such that the followings hold.\medskip

(Condition A) $\Omega_{n'} \to \Omega$ for some symmetric positive definite matrix $\Omega$.

(Condition B) $J_0[w_{n'}] = \overline J$ for all $n'$ in the subsequence.

(Condition C) As $n' \to \infty$, for each $J \subset \overline J$ and for each $j \notin \overline J'$, $|[B \tilde P_J \Omega^{-1} \mu_{n'}]_{j}| \to \infty$, and for each $j \in \overline J'$, $[B \tilde P_J \Omega^{-1} \mu_{n'}]_{j} \to a_{j,J}$ for some $a_{j,J} \in \mathbf{R}$.\medskip

In these conditions, we allow $\overline J$ or $\overline J'$ to be empty. We fix this subsequence $\{n'\}$ in Lemmas \ref{lemm: bounded sequences00}-\ref{lemm: Dn2} below. Note that $\tilde \Omega_{n'} \to_{a.s.} \Omega$ along this subsequence. Also, along this subsequence, $\tilde \Lambda^\circ\left(w_{n'},\Omega \right)$ and $\tilde \Lambda_{J}^\circ\left(w_{n'},\Omega \right)$ depend on $w_{n'}$ only through $J_0[w_{n'}] = \overline J$. Hence, we simply write $\tilde \Lambda^\circ\left(\Omega \right)$ and $\tilde \Lambda_{J}^\circ\left(\Omega \right)$ instead. Let 
\begin{align*}
    \tilde Y_{n'} := \tilde \Omega_{n'}^{1/2} \tilde Z_{n'} + \mu_{n'} \text{ and } \tilde Y_{n'}^\infty := \Omega^{1/2} \tilde Z + \mu_{n'}.
\end{align*}
For simplicity, we also let 
\begin{align*}
	\tilde \Pi_{n'} \eqdef \Pi_{\tilde \Omega_{n'}}(\tilde Y_{n'} \mid \tilde \Lambda(w_{n'})),\text{ and } \tilde \Pi_{n'}^\infty \eqdef \Pi_{\Omega}(\tilde Y_{n'}^\infty \mid \tilde \Lambda(w_{n'})),
\end{align*} 
and $\tilde R_{n'} \eqdef \tilde Y_{n'} - \tilde \Pi_{n'}$ and $\tilde R_{n'}^\infty \eqdef \tilde Y_{n'}^\infty - \tilde \Pi_{n'}^\infty.$ Clearly, $\tilde \Lambda(w_{n'})$ is a sequence of nonempty, closed and convex cones. Fix $\epsilon \in (0,1)$ and $M_\epsilon>0$ such that $\mathbf{P}(\Upsilon_{\epsilon}) > 1 - \epsilon$, where 
\begin{align}
	\label{Upsilon epsilon}
   \Upsilon_{\epsilon} = \left\{\omega \in \Upsilon': \limsup_{n' \to \infty} \left\| \tilde \Omega_{n'}(\omega)^{1/2} \tilde Z_{n'}(\omega) \right\|_{\tilde \Omega_{n'}(\omega)} < M_\epsilon\right\},
\end{align} 
where $\Upsilon' \in \mathcal{F}$ denotes the event on which $\tilde Z_{n'} \to_{a.s.} \tilde Z$ and $\tilde \Omega_{n'} \to_{a.s.} \Omega$, while $\mathbf{P}(\Upsilon') = 1$. 

\begin{lemma}
     \label{lemm: bounded sequences00}
 For each $\omega \in \Upsilon_\epsilon$, the sequences $\tilde R_{n'}(\omega)$ and $\tilde R_{n'}^\infty(\omega)$ are bounded.
\end{lemma}

\begin{proof}
   We fix $\omega \in \Upsilon_{\epsilon}$ and focus on the sequence $\tilde R_{n'}(\omega)$. The other sequence can be handled exactly in the same way. Since $\mu_{n'} \in \tilde \Lambda(w_{n'})$, by the contractivity of projection,
   \begin{align*}
	 \left\|\mu_{n'} - \tilde \Pi_{n'}(\omega) \right\|_{\tilde \Omega_{n'}(\omega)} &= \left\| \Pi_{\tilde \Omega_{n'}(\omega)}(\mu_{n'} \mid \tilde \Lambda(w_{n'})) - \Pi_{\tilde \Omega_{n'}(\omega)}(\tilde Y_{n'}(\omega) \mid \tilde \Lambda(w_{n'})) \right\|_{\tilde \Omega_{n'}(\omega)}\\
	 &\le \|\tilde \Omega_{n'}^{1/2}(\omega)\tilde Z_{n'}(\omega)\|_{\tilde \Omega_{n'}(\omega)} \le M_\epsilon.
   \end{align*}
   Note that 
   \begin{align}
	\label{proj left}
	\tilde R_{n'}(\omega) = \tilde Y_{n'}(\omega) - \tilde \Pi_{n'}(\omega)= \tilde \Omega_{n'}^{1/2}(\omega)\tilde Z_{n'}(\omega) + \mu_{n'} - \tilde \Pi_{n'}(\omega).
   \end{align}
   Since $\tilde \Omega_{n'}^{1/2}(\omega)\tilde Z_{n'}(\omega) \to \Omega^{1/2}\tilde Z(\omega)$, along the subsequence $\{n'\}$, and the sequence $\mu_{n'} - \tilde \Pi_{n'}(\omega)$ is bounded, so is $\tilde R_{n'}(\omega)$.
\end{proof}

\begin{lemma}
     \label{lemm: bounded sequences0}
 For each $\omega \in \Upsilon_\epsilon$, $\tilde R_{n'}(\omega) - \tilde R_{n'}^\infty(\omega) \to 0$, as $n' \to \infty$.   
\end{lemma}

\begin{proof}
    We write $\tilde R_{n'} - \tilde R_{n'}^\infty = A_{n',1} + A_{n,2}$, where 
    \begin{align*}
        A_{n',1} &= \Pi_{\tilde \Omega_{n'}}(\tilde Y_{n'} \mid \tilde \Lambda^\circ(\tilde \Omega_{n'})) - \Pi_{\tilde \Omega_{n'}}(\tilde Y_{n'}^\infty \mid \tilde \Lambda^\circ(\tilde \Omega_{n'})) \text{ and }\\
        A_{n',2} &= \Pi_{\tilde \Omega_{n'}}(\tilde Y_{n'}^\infty \mid \tilde \Lambda^\circ(\tilde \Omega_{n'})) - \Pi_{\Omega}(\tilde Y_{n'}^\infty \mid \tilde \Lambda^\circ(\Omega)).
    \end{align*}
  Since $\tilde \Omega_{n'}^{1/2}(\omega)\tilde Z_{n'}(\omega) \to \Omega^{1/2}\tilde Z(\omega)$ for each $\omega \in \Upsilon'$, we have $A_{n',1}(\omega) \to 0$ as $n' \to \infty$, by applying Lemma \ref{lemm: conv0}. 
  
  By Moreau decomposition,  
  \begin{align*}
    A_{n',2}(\omega) &= (\tilde Y_{n'}^\infty(\omega) - \Pi_{\tilde \Omega_{n'}(\omega)}(\tilde Y_{n'}^\infty(\omega) \mid \tilde \Lambda(w_{n'}))) - (\tilde Y_{n'}^\infty(\omega) - \Pi_{\Omega}(\tilde Y_{n'}^\infty(\omega) \mid \tilde \Lambda(w_{n'})))\\
    &= \Pi_{\Omega}(\tilde Y_{n'}^\infty(\omega) \mid \tilde \Lambda(w_{n'})) - \Pi_{\tilde \Omega_{n'}(\omega)}(\tilde Y_{n'}^\infty(\omega) \mid \tilde \Lambda(w_{n'})) \to 0,
  \end{align*}
  as $n' \to \infty$, by Lemma \ref{lemm: conv0} with Lemma \ref{lemm: bounded sequences00}.
\end{proof}

For each $\omega \in \Upsilon$, there exists unique $(J_{n'}(\omega),J_{n'}^\infty(\omega)) \in 2^{\overline J} \times 2^{\overline J}$ such that (by Lemma \ref{lemm: change of var})
\begin{align*}
     \tilde \Omega_{n'}^{-1}(\omega)\tilde R_{n'}(\omega) &= \Pi_{\tilde \Omega_{n'}^{-1}(\omega)}(\tilde \Omega_{n'}^{-1}(\omega)\tilde Y_{n'}(\omega) \mid \tilde \Lambda^\circ(I)) \in \text{ri}\left(\tilde \Lambda_{J_{n'}(\omega)}^\circ\left(I \right)\right) \text{ and } \\
     \Omega^{-1} \tilde R_{n'}^\infty(\omega) &= \Pi_{\Omega^{-1}}(\Omega^{-1}\tilde Y_{n'}^\infty(\omega) \mid \tilde \Lambda^\circ(I)) \in \text{ri}\left(\tilde \Lambda_{J_{n'}^\infty(\omega)}^\circ\left( I \right)\right).   
\end{align*}

Since $\Upsilon_\epsilon = \bigcup_{J,J' \subset \overline J} \Upsilon_{n',J,J'}$ for all $n' \ge 1$, it is not hard to see that 
\begin{align}
    \label{union}
   \Upsilon_\epsilon = \bigcup_{J,J' \subset \overline J} \Upsilon_{J,J'}. 
\end{align}

For each pair $J,J' \subset \overline J$, we let
\begin{align*}
    \Upsilon_{n',J,J'} \eqdef \left\{ \omega \in \Upsilon_\epsilon: J_{n'}(\omega) = J \text{ and } J_{n'}^\infty(\omega) = J' \right\} \text{ and } \Upsilon_{J,J'} \eqdef \bigcap_{s \ge 1} \bigcup_{n' \ge s} \Upsilon_{n',J,J'}.
\end{align*} 
We also define 
\begin{align*}
	\mathcal{J}_2(\omega) \eqdef \left\{(J,J') \in 2^{\overline J} \times 2^{\overline J}: \omega \in \Upsilon_{J,J'}\right\}.
\end{align*}

\begin{lemma}
   \label{lemm: bound232}
   For each $\omega \in \Upsilon_\epsilon$, 
\begin{align*}
    \mathcal{J}_2(\omega) = \left\{(J,J') \in 2^{\overline J'} \times 2^{\overline J'}: \omega \in \Upsilon_{J,J'}\right\}.
\end{align*}
\end{lemma}

\begin{proof}
   By Condition (C) and the choice of the subsequence $\{n'\}$, for all $n' \ge 1$ and all $\omega \in \Upsilon_\epsilon$ and $J \subset \overline J'$ and $j \notin \overline J'$, $\left|[B \tilde P_{J} \Omega^{-1} \mu_{n'}]_j \right| \to \infty$, so that
   \begin{align}
    \label{terms}
          \left|[B \tilde P_{J} \Omega^{-1} \tilde Y_{n'}^\infty(\omega)]_j \right| \to \infty.
   \end{align}
   Then, note that 
   \begin{align}
    \label{terms2}
    \left|[B \tilde P_{n',J}(\omega)\tilde \Omega_{n'}^{-1}(\omega) \tilde Y_{n'}(\omega)]_j \right|  &\ge \left|[B \tilde P_{n',J}(\omega)\tilde \Omega_{n'}^{-1}(\omega) \mu_{n'}]_j \right| - O(1)\\ \notag
    &\ge \left|[B \tilde P_{J}\Omega^{-1} \mu_{n'}]_j \right| - o\left( \| \mu_{n'} \| \right) - O(1) \to \infty,
   \end{align}
   where the inequalities follow because $\tilde P_{n',J}(\omega) \to \tilde P_J$, $\tilde \Omega_{n'}(\omega) \to \Omega$, and $\|\tilde \Omega_{n'}^{1/2}(\omega) \tilde Z_{n'}(\omega)\|_{\tilde \Omega_{n'}(\omega)} < M_\epsilon$ and due to the choice of $j \notin \overline J'$. If $\omega \in \Upsilon_{n',J,J'}$, by Lemma 3.13.2 of \cite{Silvapulle/Sen:05:ConstrainedStatInference},
   \begin{align*}
    \tilde P_{n',J}(\omega)\tilde \Omega_{n'}^{-1}(\omega) \tilde Y_{n'}(\omega) \in \text{ri}\left(\tilde \Lambda_J^\circ(I)\right) \text{ and }
    \tilde P_{J'} \Omega^{-1} \tilde Y_{n'}^\infty(\omega) \in \text{ri}\left(\tilde \Lambda_{J'}^\circ(I)\right).
   \end{align*}
   The left-hand side terms in (\ref{terms}) and (\ref{terms2}) without the absolute value diverge to $-\infty$ in this case. This means that we have $\omega \in \Upsilon_{n',J,J'}$ infinitely often only for those pairs $J,J' \subset \overline J'$.
\end{proof}

\begin{lemma}
    \label{lemm: bounded sequences}
   There exists an event $N$ with $\mathbf{P}(N) = 0$ such that for any $\omega \in \Upsilon_\epsilon \setminus N$ and $(J,J') \in \mathcal{J}_2(\omega)$, we have $J \subset J'$.
\end{lemma}

\begin{proof}
   Fix $\omega \in \Upsilon_\epsilon$ and take $(J,J') \in \mathcal{J}_2(\omega)$. Suppose that $\overline J' = \varnothing$. Then, $\mathcal{J}(\omega) = \{(\varnothing,\varnothing)\}$ by Lemma \ref{lemm: bound232}, so that $(J,J') = (\varnothing,\varnothing)$ and the lemma follows trivially. From here on, we assume that $\overline J' \ne \varnothing$.  Consider a subsequence $\{n''\} \subset \{n'\}$ such that $\omega \in \Upsilon_{n'',J,J'}$. By Lemma 3.13.2 of \cite{Silvapulle/Sen:05:ConstrainedStatInference} and Lemma \ref{lemm: change of var}, we can write 
   \begin{align*}
	  B \tilde \Omega_{n''}^{-1}(\omega) \tilde R_{n''}(\omega) &= B \tilde P_{n'',J}(\omega) \tilde \Omega_{n''}^{-1}(\omega) \tilde Y_{n''}(\omega), \text{ and }\\ 
	  \left[B \Omega^{-1} \tilde R_{n''}^\infty(\omega)\right]_{\overline J'} &= \left[B \tilde P_{J'} \Omega^{-1} \left(\Omega^{1/2} \tilde Z(\omega) + \mu_{n''} \right)\right]_{\overline J'}
      \to \left[B \tilde P_{J'} \Omega^{-1/2} \tilde Z(\omega) + a_{J'}\right]_{\overline J'},
   \end{align*}
   where $a_{J'} \in \mathbf{R}^{K+L-1}$ is a vector such that $[a_{J'}]_j = a_{j,J'}$ for each $j \in \overline J'$ and $[a_{J'}]_j = 0$ for each $j \notin \overline J'$. The last convergence follows from Conditions (A)-(C). Note that 
   \begin{align}
    \label{eq33}
     \tilde P_{n'',J}^{\perp}(\omega) B'B \tilde P_{n'',J}(\omega) \tilde \Omega_{n''}^{-1}(\omega) \tilde Y_{n''}(\omega) = \tilde P_{n'',J}^{\perp}(\omega) \tilde P_{n'',J}(\omega) \tilde \Omega_{n''}^{-1}(\omega) \tilde Y_{n''}(\omega) = 0,
   \end{align}
   where $\tilde P_{n'',J}^{\perp}$ denotes the projection matrix onto the orthogonal complement of $L_J(I)$ along $\langle \cdot,\cdot \rangle_{\tilde \Omega_{n''}^{-1}}$. The first equality follows because $B' B$ is the projection matrix onto the first $K-1$ coordinates while annihilating the last $L$ coordinates and $\tilde P_{n'',J}^\perp$ annihilates the last $L$-coordinates. Since the sequence $B \Omega^{-1} \tilde R_{n''}^\infty(\omega)$ is bounded, we have 
   \begin{align*}
    \left[\tilde P_{n'',J}^{\perp}(\omega) B' B \Omega^{-1}\tilde R_{n''}^\infty(\omega) \right]_{\overline J'}  \to \left[\tilde P_{J}^\perp \tilde P_{J'} \Omega^{-1/2} \tilde Z(\omega) + \tilde P_{J}^\perp B'  a_{J'} \right]_{\overline J'},
   \end{align*}
   where we used the convergence $\tilde P_{n'',J}^{\perp}(\omega) \to \tilde P_{J}^{\perp}$, with $\tilde P_{J}^{\perp}$ denotes the projection matrix projecting onto the orthogonal complement of $L_J(I)$ along $\langle \cdot,\cdot \rangle_{\Omega^{-1}}$. From (\ref{eq33}) and Lemma \ref{lemm: bounded sequences0}, we find that 
   \begin{align*}
      \left[\tilde P_{J}^\perp B' B \tilde P_{J'} \Omega^{-1/2} \tilde Z(\omega) + \tilde P_{J}^\perp B'  a_{J'} \right]_{\overline J'}= \left[\tilde P_{J}^\perp \tilde P_{J'} \Omega^{-1/2} \tilde Z(\omega) + \tilde P_{J}^\perp B'  a_{J'} \right]_{\overline J'} = 0.
   \end{align*}
   We let $N_{J,J'}$ be the collection of $\omega \in \Upsilon$ such that the last equality holds and let 
   \begin{align*}
	  \mathcal{J}_2^\circ \eqdef \left\{(J,J') \in 2^{\overline J} \times 2^{\overline J}: \tilde P_{J}^\perp \tilde P_{J'} = 0 \right\}.
   \end{align*} Then, we have shown that  
   \begin{align*}
	   \Upsilon_\epsilon \subset N \cup \tilde N,
   \end{align*}
   where 
   \begin{align*}
	 N \eqdef \bigcup_{(J,J') \in (2^{\overline J} \times 2^{\overline J}) \setminus \mathcal{J}_2^\circ} N_{J,J'} \text{ and } \tilde N \eqdef \bigcup_{(J,J') \in \mathcal{J}_2^\circ} N_{J,J'}.
   \end{align*}
   Since $\Omega^{-1/2} \tilde Z$ is a continuous random vector, if $\tilde P_J^\perp \tilde P_{J'} \ne 0$, then, we must have $\mathbf{P}(N_{J,J'}) = 0$. Hence, $\mathbf{P}(N) = 0$ and
   \begin{align*}
	   \Upsilon_\epsilon \setminus N \subset \tilde N.
   \end{align*}  
   That is, for each $\omega \in \Upsilon_\epsilon \setminus N$ and $(J,J') \in \mathcal{J}_2(\omega)$, we must have $\tilde P_J^\perp \tilde P_{J'} = 0$. Since $|J \cup J'| \le |\overline J| \le K-1$, in light of Lemma \ref{lemm: rank}, this implies that $J \subset J'$.
\end{proof}

\begin{lemma}
	\label{lemm: conv}
	Let $N$ be the null event in Lemma \ref{lemm: bounded sequences}. Then, for all $\omega \in \Upsilon_\epsilon \setminus N$, there exists a subsequence $\{n''\} \subset \{n'\}$,
    \begin{align}
        \label{limit J}
        J_0\left[ B \tilde \Omega_{n''}^{-1}(\omega) \tilde R_{n''}(\omega)\right] \cap \overline J \subset J_0\left[B \Omega^{-1} \tilde R_{n''}^\infty(\omega) \right] \cap \overline J.
    \end{align}
\end{lemma}

\begin{proof}
    Fix $\omega \in \Upsilon_\epsilon \setminus N$ and take $(J,J') \in \mathcal{J}_2(\omega)$. Choose a subsequence $\{n''\}$ such that $\omega \in \Upsilon_{n'',J,J'}$ for all $n''$ in the subsequence. By Lemma \ref{lemm: partition}, the set on the left-hand side of (\ref{limit J}) is $J$ and that on the right-hand side is $J'$. The desired result follows from Lemma \ref{lemm: bounded sequences} and (\ref{union}).
\end{proof}

   For each $\epsilon$, we let the event $D_n(\epsilon)$ be that of
    \begin{align}
	   \label{ineq340}
     1\left\{\left\|\tilde R_{n} \right\|_{\tilde \Omega_{n}}^2 > c_{1-\alpha,\tilde \Omega_{n}}(\tilde Y_{n};w_{n}) \right\} \le 1\left\{\left\|\tilde R_{n}^\infty  \right\|_{\Omega}^2 > c_{1-\alpha,\Omega}(\tilde Y_{n}^\infty;w_{n}) - \epsilon \right\}.
    \end{align}

    \begin{lemma}
        \label{lemm: Dn}
       Let $N$ be the null event in Lemma \ref{lemm: bounded sequences}. Then, $\Upsilon_{\epsilon} \setminus N \subset \liminf_{n \to \infty} D_{n}(\epsilon).$
    \end{lemma}
    
    \begin{proof}
        We fix $\omega \in \Upsilon_\epsilon \setminus N$. We let $\{n''\}$ be the subsequence in Lemma \ref{lemm: conv}. Since $\tilde \Omega_{n''}(\omega) = \Omega + o(1)$ and, by Lemma \ref{lemm: bounded sequences0}, $\tilde R_{n''}^\infty(\omega) = \tilde R_{n''}(\omega) + o(1)$, as $n'' \to \infty$,
        \begin{align*}
            \left\|\tilde R_{n''}(\omega) \right\|_{\tilde \Omega_{n''}(\omega)}^2 = \left\|\tilde R_{n''}^\infty(\omega) \right\|_{\Omega}^2 + o(1),
        \end{align*}
        because $\tilde R_{n''}(\omega)$ and $\tilde R_{n''}^\infty(\omega)$ are bounded by Lemma \ref{lemm: bounded sequences00}. 
        
        On the other hand, note that 
        \begin{align*}
            B \tilde \Omega_{n''}^{-1}(\omega) \tilde R_{n''}(\omega) = \pi_{\tilde \Omega_{n''}(\omega)}^\circ(\tilde Y_{n''}(\omega) ; w_{n''}) \text{ and } B \Omega^{-1} \tilde R_{n''}^\infty(\omega) = \pi_{\Omega}^\circ(\tilde Y_{n''}^\infty(\omega) ; w_{n''}).
        \end{align*}
         Hence, by Lemma \ref{lemm: conv},
\begin{align*}
    \left|J_0\left[\pi_{\tilde \Omega_{n''}(\omega)}^\circ(\tilde Y_{n''}(\omega) ; w_{n''})\right] \cap J_0[w_{n''}] \right| \le \left| J_0\left[\pi_{\Omega}^\circ(\tilde Y_{n''}^\infty(\omega) ; w_{n''})\right] \cap J_0[w_{n''}] \right|,
\end{align*}
which implies that
\begin{align*}
	c_{1-\alpha,\tilde \Omega_{n''}(\omega)}\left(\tilde Y_{n''}(\omega);w_{n''}\right) \ge c_{1-\alpha,\Omega}\left(\tilde Y_{n''}^\infty(\omega);w_{n'}\right).
\end{align*}
Thus, from some large $n''$ on,
\begin{align}
    \label{ineq2}
    & 1\left\{\left\| \tilde Y_{n''}(\omega) - \tilde \Pi_{n''}(\omega) \right\|_{\tilde \Omega_{n''}(\omega)}^2 > c_{1-\alpha,\tilde \Omega_{n''}(\omega)}(\tilde Y_{n''}(\omega);w_{n''}) \right\}\\ \notag
    &\le 1\left\{\left\| \tilde Y_{n''}^\infty(\omega) - \tilde \Pi_{n''}^\infty(\omega) \right\|_{\Omega}^2 > c_{1-\alpha,\Omega}(\tilde Y_{n''}^\infty(\omega);w_{n''}) - \epsilon \right\}. \notag
\end{align} 
That is, $\omega \in D_{n''}(\epsilon)$ from some large $n''$ on. The desired result follows because $\{n''\}$ is a subsequence of $\{n'\}$ which is a subsequence of any arbitrary subsequence of $\{n\}$. 
\end{proof}

\begin{lemma}
    \label{lemm: Dn2}
    $\liminf_{n \to \infty} \mathbf{P}(D_{n}(\epsilon)) > 1 - \epsilon.$
\end{lemma}

\begin{proof}
    By Lemma \ref{lemm: Dn}, 
    \begin{align*}
        \liminf_{n \to \infty} \mathbf{P}(D_{n}(\epsilon)) &\ge \liminf_{n \to \infty} \mathbf{P}(D_{n}(\epsilon) \cap \Upsilon_{\epsilon}) \ge \mathbf{P}\left( \liminf_{n \to \infty} D_{n}(\epsilon) \cap \Upsilon_{\epsilon}\right) = \mathbf{P}(\Upsilon_\epsilon) > 1 - \epsilon.
    \end{align*}
\end{proof}

\noindent \textbf{Proof of Lemma \ref{lemm: Mohamad2}: } By Lemma \ref{lemm: bounded sequences0},
\begin{align}
    \label{conv343}
	\left\| \tilde Y_{n} - \tilde \Pi_{n} \right\|_{\tilde \Omega_{n}}^2 - \left\|\tilde Y_{n}^\infty - \tilde \Pi_n^\infty \right\|_{\Omega}^2 \rightarrow_{a.s.} 0, \text{ as } n \to \infty.
\end{align}

Let us turn to the critical values. Note that there exists a continuous map $\eta: [0,\infty) \rightarrow [0,\infty)$ such that $\lim_{x \rightarrow 0} \eta(x) = 0$ and for any $t > 0$ such that $0< \eta(t)< 1-\alpha$,
\begin{align*}
 G^{-1}\left(1-\alpha - \eta(t); d\right)  \le G^{-1}\left(1-\alpha ; d\right) - t, \text{ for all } 1\le d \le K + L -1.
\end{align*} 
Fix $\epsilon \in (0,1)$ such that $0< \eta(\epsilon)< \alpha$. Then,
\begin{align}
	\label{ineqs23}
	&\limsup_{n \to \infty} \mathbf{P}\left\{\left\|\tilde Y_{n} - \tilde \Pi_{n} \right\|_{\tilde \Omega_{n}}^2 > c_{1-\alpha,\tilde \Omega_{n}}(\tilde Y_{n};w_{n})\right\}\\ \notag
	&\le \limsup_{n \to \infty}\mathbf{P}\left(\left\{\left\|\tilde Y_{n} - \tilde \Pi_{n} \right\|_{\tilde \Omega_{n}}^2 > c_{1-\alpha,\tilde \Omega_{n}}(\tilde Y_{n};w_{n}) \right\} \cap D_{n}(\epsilon)\right) + \limsup_{n \to \infty}\mathbf{P}(D_{n}^c(\epsilon))\\ \notag
    &\le \limsup_{n \to \infty}\mathbf{P}\left(\left\{\left\|\tilde Y_{n}^\infty - \tilde \Pi_{n}^\infty \right\|_{\Omega}^2 > c_{1-\alpha,\Omega}(\tilde Y_{n}^\infty;w_{n}) - \epsilon \right\} \cap D_{n}(\epsilon)\right) + \epsilon\\ \notag
	&\le \limsup_{n \to \infty}\mathbf{P}\left(\left\{\left\|\tilde Y_{n}^\infty - \tilde \Pi_{n}^\infty \right\|_{\Omega}^2 > c_{1-\alpha -\eta(\epsilon),\Omega}(\tilde Y_{n}^\infty;w_{n}) \right\} \cap D_{n}(\epsilon)\right) + \epsilon  \le \alpha + \eta(\epsilon) + \epsilon,
\end{align}
where the second inequality follows by Lemma \ref{lemm: Dn2} and the last inequality by Lemma \ref{lemm: Mohamad}. By sending $\epsilon \to 0$, we complete the proof. $\blacksquare$\medskip

\noindent \textbf{Proof of Theorem \ref{thm: main}: } First, note that for $(w,\theta) \in \Delta_{K-1} \times \Theta$,
\begin{align*}
	\hat d(w,\theta) &= \left|J_0\left[B \hat \Omega^{-1}(w) \tilde B_2' \hat s(w,\theta,\hat \lambda(w))\right] \cap J_0[w] \right|\\
	&= \left|J_0\left[B \hat \Omega^{-1}(w) (Y_n(w,\theta) - \Pi_{\hat \Omega(w)}(Y_n(w,\theta) \mid \tilde \Lambda(w)))\right] \cap J_0[w] \right|,
\end{align*}
where $Y_n(w,\theta)$ is defined in (\ref{Yn(w)}).

Since $x = \Pi_{\Omega}(x \mid \tilde \Lambda(w)) + \Pi_{\Omega}(x \mid \tilde \Lambda^\circ(w,\Omega))$ for any $x \in \mathbf{R}^{K + L -1}$, the last term is equal to 
\begin{align*}
	\left|J_0\left[B \hat \Omega^{-1}(w) \Pi_{\hat \Omega(w)}(Y_n(w,\theta) \mid \tilde \Lambda^\circ(w,\hat \Omega(w)))\right]\cap J_0[w] \right|.
\end{align*}
Using the definition of $c_{1-\alpha, \Omega}(\cdot;w)$ in (\ref{eq: critical value}), we can write $\hat c_{1-\alpha}(w,\theta)$ in Algorithm \ref{alg:confidence_set3} as
\begin{align*}
	\hat c_{1-\alpha}(w,\theta) = c_{1-\alpha, \hat \Omega(w)}(Y_n(w,\theta);w).
\end{align*}
It suffices to show the following: along any sequence $P_n \in \mathcal{P}_n$, 
\begin{align*}
	\lim_{n \rightarrow \infty} \sup_{w \in \mathbb{W}_{P_n}} P_n\left\{T(w,\theta_{P_n}(w)) > c_{1-\alpha, \hat \Omega(w)}\left(Y_n(w,\theta_{P_n}(w));w \right) \right\} \le \alpha.
\end{align*}
By the definition of the supremum, there exists a sequence $w_n \in \mathbb{W}_{P_n}$ such that the limit on the left hand side is equal to 
\begin{align}
	\label{eq: validity}
	\lim_{n \rightarrow \infty} P_n\left\{T(w_n,\theta_{P_n}(w_n)) > c_{1-\alpha, \hat \Omega(w_n)}\left(Y_n(w_n,\theta_{P_n}(w_n));w_n \right) \right\}.
\end{align}
We apply Lemma \ref{lemm: Mohamad2} by setting
\begin{align*}
    Y_n = Y_n(w_n,\theta_{P_n}(w_n)), \enspace Z_n = \hat \Omega^{-1/2} (Y_n - \mu_n), \enspace \Omega_n = \Omega_{P_n}(w_n), \text{ and } \mu_n = \mu_n(w_n),
\end{align*}
where $\hat \Omega = \hat \Omega(w_n)$ and $\mu_n(w_n)$ is defined in (\ref{Zn(w)}). By Assumption \ref{assump: asym normal}, Assumption \ref{basic assumption} is satisfied. From the Karush-Kuhn-Tucker condition (\ref{equality restr}), we have $\mu_n(w_n) \in \tilde \Lambda(w_n)$.  Thus, the limit in (\ref{eq: validity}) is bounded by $\alpha$ by Lemma \ref{lemm: Mohamad2}. $\blacksquare$

\end{document}